%% file: main_arxiv.tex
\documentclass[11pt,letterpaper]{article}
\usepackage[left=0.85in,right=0.85in,top=0.85in,bottom=0.9in]{geometry}
\usepackage[T1]{fontenc}
\usepackage[utf8]{inputenc}
\usepackage{amsmath,amssymb}
\usepackage{graphicx}
\usepackage{booktabs}
\usepackage{array}
\usepackage{longtable}
\usepackage{pdflscape}
\usepackage{ragged2e}
\usepackage{tabularx}
\usepackage{multirow}
\usepackage{caption}
\usepackage{subcaption}
\usepackage{float}
\usepackage[numbers,sort&compress]{natbib}
\usepackage{microtype}
\usepackage{hyperref}
\hypersetup{colorlinks=true,linkcolor=blue,citecolor=blue,urlcolor=blue}
\graphicspath{{./figures/}}
\renewcommand{\arraystretch}{1.12}
\newcommand{\cmvs}{cm$^2$ V$^{-1}$ s$^{-1}$}

\begin{document}

\title{InvDesMobility: a reliability-gated first-principles feedback framework for closed-loop materials discovery}
\author{Wen-Kao Li$^{1,\dagger}$, Ze-Feng Gao$^{1,\dagger,*}$, Peng-Jie Guo$^1$, Wei Ji$^1$,  Zhong-Yi Lu$^{1,*}$\\
\small $^1$School of Physics and Key Laboratory of Quantum State Construction \\
\small and Manipulation (Ministry of Education), Renmin University of China, Beijing 100872, China\\
\small $^\dagger$These authors contributed equally.\\
\small $^*$e-mail: zfgao@ruc.edu.cn; zlu@ruc.edu.cn}
\date{}
\maketitle
\vspace{-1.25em}

\begin{abstract}
Inverse materials design starts from target functionality and searches for structures that can realize it. Its value in closed-loop discovery depends not only on prediction performance, but also on whether expensive first-principles results are independently validated, provenance-recorded, and admitted as feedback only when evidence is sufficient. This is especially important for composite properties such as carrier mobility, where a final scalar value hides intermediate quantities, fit quality, convergence history, and workflow assumptions. Here we present \textit{InvDesMobility}, a reliability-gated first-principles feedback framework that integrates multi-agent automated DFT, evidence stratification, generative structure proposal, acquisition ranking, and auditable release. Using 516 2DMatPedia-derived candidates, the workflow produced 280 QC-passed materials and 573 retained carrier--direction seed channels after channel-level reliability gating. These records were split into two feedback objects: relaxed structures updated the generative model, while retained mobility channels trained the acquisition model and set validation priority. Over multiple iterations, \textit{InvDesMobility} screened $2.4\times10^6$ structures, submitted 102 candidates for DFT validation, and retained 86 reliability-gated generated channels across 41 formulas. Overall, the main contribution is not a fixed list of high-mobility materials, but a transferable feedback contract that makes closed-loop inverse design both useful and auditable when learning from expensive calculated properties. All source data, retained feedback records, and workflows are available at \url{https://github.com/DreamLufei/invDesMobility}, with an accompanying evidence website at \url{https://dreamlufei.github.io/invDesMobility/}.
\end{abstract}

\noindent\textbf{Keywords:} closed-loop materials discovery, first-principles calculations, reliability-gated feedback, materials informatics, two-dimensional semiconductors, carrier mobility

\section*{Introduction}

Computational materials discovery is moving from static enumeration toward inverse design, in which a target functionality is specified first and algorithms search for compositions or structures that could realize it~\cite{Butler2018ML,Schmidt2019MLSolidState,InvDesFlow2024,InvDesFlowAL2025}. This change matters because conventional high-throughput discovery is usually forward: one starts from a database, calculates properties and then ranks the survivors. Forward screening is valuable, but it is limited by the contents of the initial library and by the cost of evaluating every candidate with density functional theory (DFT), let alone with property-specific first-principles workflows~\cite{Jain2013MP,Mounet2018Exfoliation,Haastrup2018C2DB}. Inverse design changes the direction of the problem. It uses learned structure--property relations and generative models to propose candidates beyond the original database, then asks which proposals deserve expensive validation~\cite{Xie2018CGCNN,Choudhary2021ALIGNN,DiffCSP2023,Merchant2023GNoME,Zeni2025MatterGen}.

Closed-loop materials design is the natural reliability mechanism for this inverse-design setting~\cite{Lookman2019ActiveLearning,Hase2019SDL,MacLeod2020SDL,Szymanski2023ALab}. A purely generative or predictive inverse-design workflow can extrapolate into chemically fragile regions, whereas a closed loop repeatedly returns selected candidates to a validation engine and uses the resulting evidence to reshape the next search. This matters for efficiency because scarce first-principles calculations are concentrated on candidates that can most improve the search, and it matters for reliability because the loop can correct model errors as validated evidence accumulates. The reliability is not automatic: it comes from keeping prediction, validation and feedback as separate decisions. Models may propose structures and rank validation queues, but only validated records with sufficient physical and numerical support should update the learning state. Without that separation, errors in a small number of expensive labels can be amplified into the next generation of candidates.

Generative models, graph neural networks and active-learning strategies can therefore be powerful accelerators: they can produce new candidates, rank a validation queue and concentrate scarce calculations on regions that appear promising~\cite{Xie2018CGCNN,Choudhary2021ALIGNN,Lookman2019ActiveLearning,DiffCSP2023,Merchant2023GNoME}. Yet the same acceleration creates a second, less visible bottleneck. When a loop learns from every completed calculation, the closed-loop form alone does not guarantee scientific reliability. The central question is therefore not only how to search a larger space, but how to decide which expensive calculated results are reliable enough to become feedback.

This question is especially important for first-principles properties that are not primitive outputs of a single calculation. In many materials-design tasks, the useful learning object is assembled from a chain of intermediate physical quantities: relaxed structures, band edges, curvatures, response slopes, elastic constants, convergence states, fitted uncertainties and assumptions about the physical model. A final scalar value can look complete even when part of this chain is poorly conditioned. If such a value is admitted into a generative or active-learning loop, the loop may learn from workflow artifacts rather than from physically meaningful evidence. The problem is not solved by recording provenance alone, because provenance says what happened whereas feedback control must decide what is allowed to teach the next round~\cite{Jain2015FireWorks,Huber2020AiiDA,Wilkinson2016FAIR}. A closed-loop framework therefore needs an explicit evidence standard that separates numerical completion from feedback admissibility before any model is updated.

Carrier mobility in two-dimensional (2D) semiconductors provides a stringent validation scenario for this broader problem. High mobility remains a central requirement for low-power electronics, and atomically thin semiconductors offer material-level control over electrostatics, anisotropic transport and channel geometry~\cite{Wang2012TMD,Akinwande2014,Fiori2014,Novoselov2016VdW,Chhowalla2016TwoDSemiconductors}. At the same time, mobility is a difficult property to use as closed-loop feedback. A reported channel mobility is not a direct DFT observable; it combines band-edge identity, effective masses, strain-dependent band-edge shifts, two-dimensional elastic response and a scattering model. Each of these ingredients can fail or become ambiguous in a different way. A band edge may switch under strain, an effective-mass fit may be unstable, a deformation-potential slope may be dominated by poorly conditioned points, or a scattering assumption may be used outside its range of validity~\cite{Bardeen1950,Qiao2014BP,Ponce2020,Giustino2017ElectronPhonon,Zhang2023Mobility}. Such failures can still yield finite numbers, which makes mobility an effective stress test for any framework that claims to learn from expensive first-principles properties.

The practical challenge is that mobility feedback is channel-resolved. In a 2D semiconductor, electron and hole transport along two in-plane directions can have different band extrema, effective masses, elastic responses and fitting reliability. A material can therefore have one carrier--direction channel that is reliable enough for learning while another channel is cautionary or should be withheld. Treating the whole material as a single label discards this structure; treating every finite channel as equally valid contaminates the feedback set. A useful closed-loop system must preserve the channel-level evidence, retain only the channels that satisfy the reliability rule and still keep the non-retained outputs visible for audit. This requirement makes mobility more than an application example: it exposes the general need for reliability-gated feedback in composite-property discovery.

Several pieces of the required ecosystem already exist. Databases and high-throughput infrastructures provide reference pools for crystalline and 2D materials~\cite{Jain2013MP,Mounet2018Exfoliation,Haastrup2018C2DB}. Graph models and active-learning methods can prioritize expensive calculations~\cite{Butler2018ML,Schmidt2019MLSolidState,Xie2018CGCNN,Choudhary2021ALIGNN,Lookman2019ActiveLearning}. Autonomous laboratories demonstrate how computation, decision-making and validation can be coupled into iterative campaigns~\cite{Szymanski2023ALab,Hase2019SDL,MacLeod2020SDL}. Generative models now move materials design beyond ranking known candidates toward proposing new structures, including large-scale inorganic discovery and 2D-material-specific inverse design~\cite{DiffCSP2023,Merchant2023GNoME,Zeni2025MatterGen,Lyngby2022TwoDGenerative,Fung2021MatDesINNe,InvDesFlow2024,InvDesFlowAL2025}. These advances make closed-loop materials discovery technically feasible, but they do not by themselves define the state variable of the loop. A generated structure, a prediction score, a converged calculation and a retained property record are different objects. Prediction can nominate validation work; it cannot certify that a completed property value should become a training signal.

We frame the missing abstraction as feedback admissibility. Feedback admissibility is a pre-learning decision about whether a completed property record can update generation, acquisition or subsequent model selection. It is stricter than convergence and more operational than retrospective data cleaning: it must assign completed outputs to retained feedback, cautionary evidence or withheld records before any result can influence the next round. Workflow systems can record provenance~\cite{Jain2015FireWorks,Huber2020AiiDA}, and model validation can summarize performance after a campaign, but neither step alone prevents unstable intermediate quantities from reshaping the search while the campaign is still running. A general closed-loop framework therefore needs a feedback contract that connects automated calculation, reliability assessment, record retention and learning-state updates in one auditable sequence.

\textit{InvDesMobility} implements this contract as a reliability-gated first-principles feedback framework. The framework separates three objects that are often conflated in inverse design. First, a proposed candidate is not a validated candidate. Second, a completed numerical output is not automatically retained feedback. Third, structure feedback and property feedback update different parts of the loop. In the implementation used here, a multi-agent automated DFT engine plans and executes the mobility workflow, tracks recovery and provenance, and produces structured carrier--direction records. A reliability gate then evaluates whether each record has the physical and numerical support required for feedback. Quality-controlled relaxed structures can adapt the generative model, whereas retained mobility channels train the acquisition model and define which generated candidates are worth validating next. Non-retained outputs remain part of the released audit trail, but they do not teach the loop.

We validate this general framework using 2D semiconductor mobility as the experimental scenario. The campaign first constructs a seed evidence base from 516 2DMatPedia-derived candidates. These materials are not merely screened by a final mobility threshold; they are evaluated by a multi-agent automated DFT mobility engine that records the intermediate evidence needed to interpret every carrier--direction channel. After detailed reliability gating, the seed stage yields 280 QC-passed materials and 573 retained carrier--direction channels. These 280 QC-passed relaxed structures are used to adapt the generative structure model, while the retained channel records define the mobility evidence used for acquisition. Thus, the seed stage creates both the structural experience needed for generation and the property evidence needed for prioritization, without treating every finite value as a learning label.

The second stage closes the loop. Generated candidates are first filtered by structural, electronic, energetic, dynamical and model-based criteria, producing a DFT-scale queue from a much larger generated space. The selected materials are then returned to the same multi-agent DFT mobility engine for automated carrier-mobility calculation, reliability assessment and feedback assignment. Only channels that pass the reliability gate are admitted as new feedback; other completed outputs remain traceable but do not update the learning state. Across multiple iterations, this process screened $2.4\times10^6$ generated structures, submitted 102 candidates to DFT validation and retained 86 reliability-gated generated channels across 41 formulas. The resulting study therefore uses mobility to validate a broader framework: closed-loop discovery should be measured not only by how many structures are proposed, but by how much reliable first-principles evidence is admitted back into the loop.

\section*{Results}

The Results present the framework first and the mobility campaign as its validation. We first establish the multi-agent DFT engine and evidence hierarchy that convert expensive calculations into admissible or non-admissible feedback records, then instantiate the framework for 2D carrier mobility. The section is organized around the decisions a general closed-loop framework must make: how seed calculations become learning-ready evidence, how retained evidence is used without treating a surrogate model as a substitute for DFT, how a very large generated space is compressed to a validation-scale queue and how generated candidates are admitted back into the feedback set only after revalidation. The seed, acquisition, screening and generated-validation analyses therefore test how much evidence survives the reliability gate, whether retained evidence can enrich a finite validation queue and which generated channels remain after first-principles revalidation. This organization keeps the central claim falsifiable: generators and surrogate models can nominate work, but only reliability-gated first-principles channels can enlarge the feedback set.

\subsection*{A multi-agent DFT engine defines admissible first-principles feedback}

\begin{figure}[htbp]
\centering
\includegraphics[width=\textwidth]{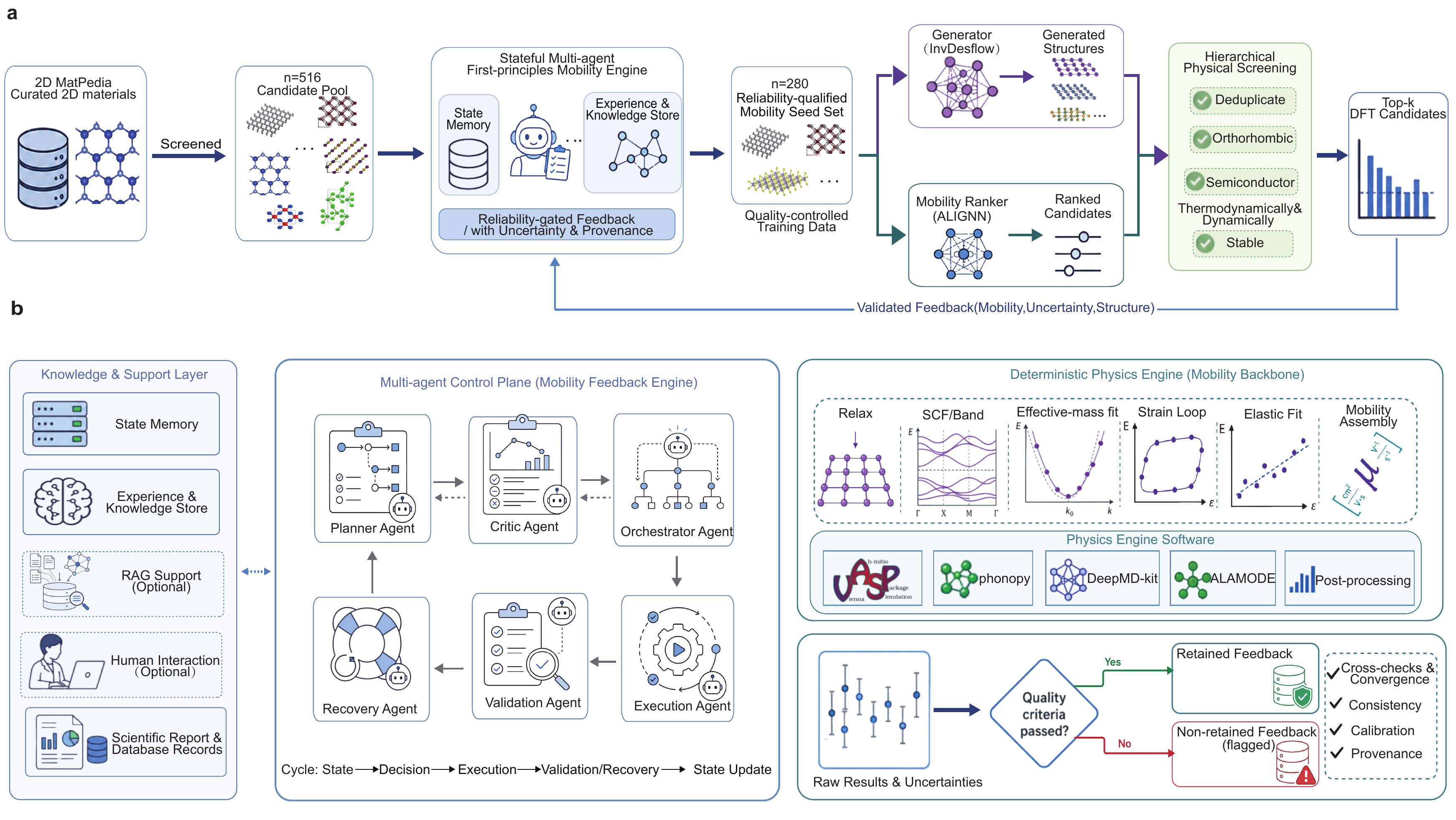}
\caption{\textbf{A reliability-gated first-principles engine turns expensive calculations into admissible feedback.} \textbf{a,} General closed-loop feedback framework instantiated for 2D carrier mobility. A 2DMatPedia-derived candidate pool is evaluated by a stateful multi-agent first-principles engine to form a quality-controlled mobility seed set. Quality-controlled relaxed structures and retained channel entries are routed separately to the generator and acquisition ranker. Generated structures are physically screened and ranked into a DFT validation queue, and only reliability-gated evidence is returned to the runtime and training sets. \textbf{b,} Internal validation architecture. A knowledge and support layer supplies state memory, retrieval context, experience records and reports to a multi-agent orchestration layer, which coordinates the VASP-based mobility backbone. The evidence hierarchy separates workflow records into retained feedback, cautionary evidence or withheld output before any property record can update the loop.}
\label{fig:framework}
\end{figure}

The first requirement for a transferable closed-loop framework is an evidence rule that operates before learning, not after a campaign has ended. \textit{InvDesMobility} implements this rule by connecting seed evidence construction, structure generation, physical screening, acquisition ranking, first-principles validation and feedback assignment into one design loop (Fig.~\ref{fig:framework}a). In the mobility validation scenario, seed and generated 2D semiconductors are evaluated by the same stateful mobility engine, so a calculation can guide the next round only after it becomes a channel-level feedback record with an assigned reliability status. Quality-controlled relaxed structures update the DiffCSP/InvDesFlow generator, retained channel-resolved mobility entries train the ALIGNN acquisition ranker, and generated structures are screened and ranked into a DFT-scale validation queue.

The internal architecture makes the feedback gate explicit (Fig.~\ref{fig:framework}b). A multi-agent orchestration layer coordinates the VASP-based mobility backbone across planning, execution, validation and recovery, while the runtime records the intermediate evidence needed to interpret a final mobility value. A finite mobility value is therefore not sufficient by itself: the loop also records the strain points that completed, the fit quality of the deformation-potential and elastic response, the effective-mass fit, the dynamic band-edge behaviour and the channel status assigned by the reliability gate. These records allow the same numerical workflow to return a retained channel for learning, cautionary evidence for audit or a withheld channel that should not update the acquisition model.

Because feedback status is assigned during execution, reproducibility becomes part of the campaign rather than an after-the-fact reconstruction. Each completed calculation is converted into retained feedback, cautionary evidence or withheld output before it can influence the next round. Table~\ref{tab:runtime_roles} summarizes the resulting guardrails, which turn workflow provenance into a feedback decision; the corresponding implementation modules, runtime policy and context-loading design are detailed in Methods, Supplementary Table~S1, Supplementary Notes~2--4 and Supplementary Fig.~S2.

\begin{table}[htbp]
\centering
\caption{\textbf{Runtime roles linking first-principles execution to feedback admissibility.} The table complements Fig.~\ref{fig:framework}b by showing how runtime components connect workflow control, provenance and retained-feedback assignment in the reliability-gated framework.}
\label{tab:runtime_roles}
\small
\renewcommand{\arraystretch}{1.16}
\begin{tabularx}{\textwidth}{@{}p{0.20\textwidth}X X@{}}
\toprule
Layer & Runtime role & Feedback role \\
\midrule
Planner--critic--orchestrator & Select workflow stages from typed state and stage contracts. & Maintains a stage-resolved execution history for provenance. \\
Execution and recovery & Launch VASP-native tools and route controlled retries or restarts. & Records completed, failed and recovered workflow states. \\
Validation gate & Check convergence, fit quality, uncertainty and channel consistency. & Assigns retained, cautionary, weak or failed channel status. \\
RAG and skills & Provide stage-scoped policy evidence and recovery context. & Supports VASP-input decisions, validation summaries and reports. \\
Feedback assignment & Separate relaxed-structure feedback from retained mobility feedback. & Routes quality-controlled structures and retained channels to the generator and ranker. \\
\bottomrule
\end{tabularx}
\end{table}

\subsection*{A mobility validation scenario converts 516 seed candidates into learning-ready feedback}

The seed campaign tests how severely a reliability rule reshapes an apparently complete mobility screen. This step is deliberately placed before generation, because a closed loop needs an initial evidence base that is already separated into feedback and non-feedback records. Starting from 516 two-dimensional semiconductor candidates from 2DMatPedia\cite{Zhou2019TwoDMatPedia}, the multi-agent DFT workflow evaluated four in-plane transport channels per material and retained 573 carrier--direction channels across 280 QC-passed materials as the seed evidence for closed-loop learning. The result is a transport-feedback database rather than a catalogue of finite mobilities: each channel is assessed for workflow completion, numerical validity, deformation-potential and elastic-fit quality, uncertainty, effective-mass reliability and band-edge consistency before being retained as feedback (Supplementary Note~1 and Supplementary Tables~S3 and S4). In this role, the seed campaign defines the framework's initial contract: every later generated candidate must satisfy the same type of evidence requirement before it can enlarge the learning set.

The resulting resource is not just a screened materials list, but a first-principles transport-feedback database. In the deformation-potential protocol, a retained channel depends on band-edge identity, effective masses, strain-dependent band-edge shifts, two-dimensional elastic response and fitted deformation potentials\cite{Bardeen1950,Ponce2020,Giustino2017ElectronPhonon,Zhang2023Mobility}. A finite mobility value can still be unsuitable for learning if the band edge switches under strain or the fits are poorly constrained. The database therefore stores both the value and the evidence that determines whether the value can become feedback.

\begin{figure}[htbp]
\centering
\includegraphics[width=0.98\textwidth]{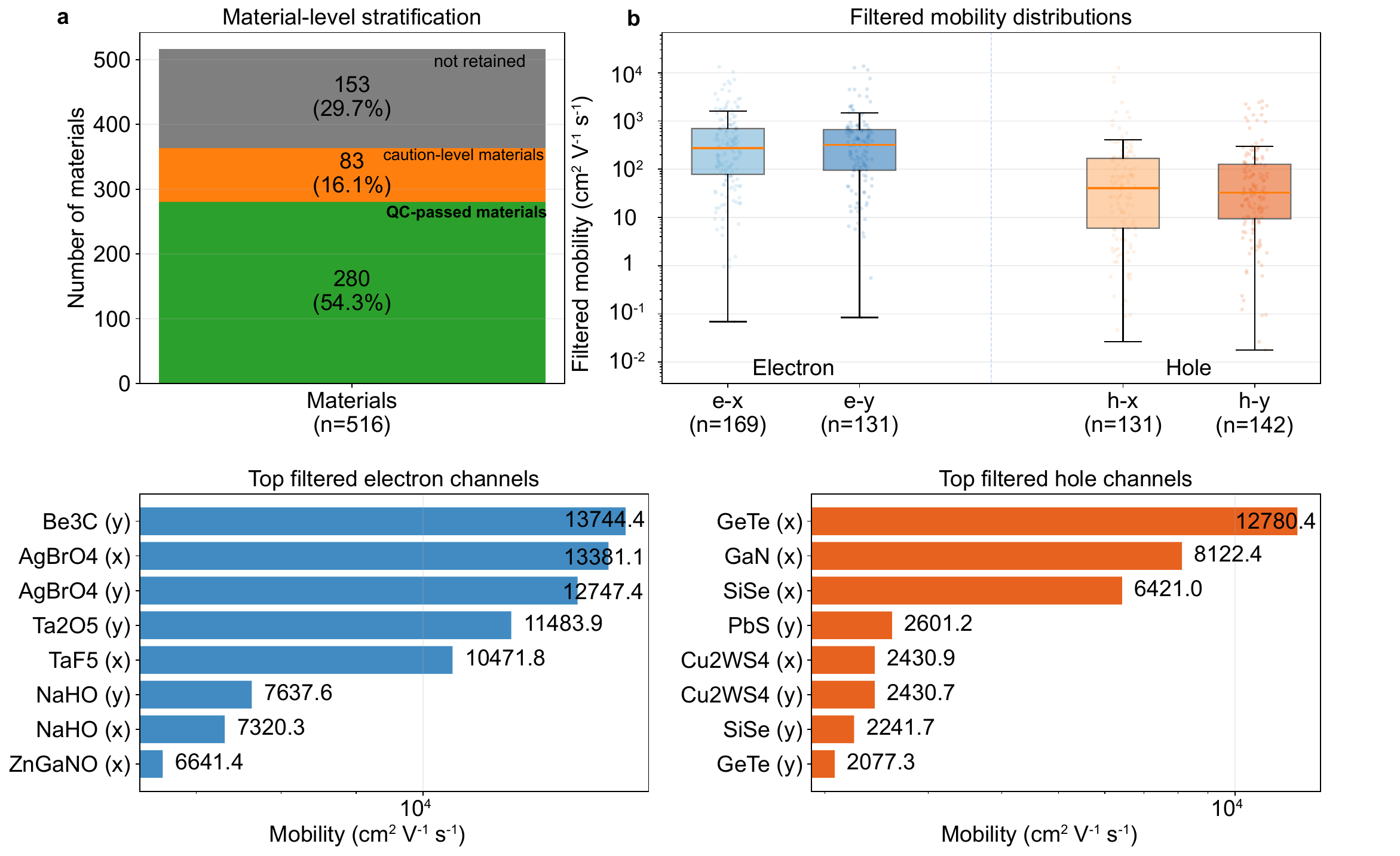}
\caption{\textbf{The mobility validation scenario converts seed calculations into learning-ready feedback.} 
\textbf{a,} Material-level reliability stratification of the 516 2DMatPedia-derived semiconductors evaluated by the InvDesMobility mobility engine. 
\textbf{b,} Retained carrier--direction mobility entries in the seed database, grouped by carrier type and in-plane transport direction. 
\textbf{c,d,} Representative high-mobility retained electron and hole channels after channel-level reliability filtering. 
The notation e-$x$, e-$y$, h-$x$ and h-$y$ denotes electron-$x$, electron-$y$, hole-$x$ and hole-$y$ channels; $n$ gives the number of retained entries in each class. Mobilities are reported in \cmvs.}
\label{fig:seed}
\end{figure}

Channel outcomes define the material-level classes shown in Fig.~\ref{fig:seed}a. A material is QC-passed if at least one channel is retained, caution-level if it has caution-level but no retained channels, and not retained otherwise. This rule yields 280 QC-passed materials, 83 caution-level records and 153 not-retained records among the 516 seed structures. Thus, QC-passed does not mean high mobility; it means that at least one carrier--direction channel is reliable enough to enter the main analysis and feedback set. This distinction is essential for anisotropic two-dimensional semiconductors, where different carrier--direction channels can have different effective masses, deformation-potential slopes and fit stability\cite{Qiao2014BP,Zhang2023Mobility}.

The retained part of the database contains 573 channel-resolved mobility entries, divided into 169 electron-$x$, 131 electron-$y$, 131 hole-$x$ and 142 hole-$y$ channels (Fig.~\ref{fig:seed}b). The non-retained evidence is also kept in the audit records: 301 seed channels are caution-level, 1,020 are weak and 170 are failed or non-usable. The remaining records are therefore not hidden; they define the boundary between numerical output and feedback admitted to the learning loop. This database-level distinction is what prevents large but physically fragile mobility values from being silently promoted into training data.

The retained seed entries span several orders of magnitude, motivating logarithmic statistics and the material-level acquisition target $\log_{10}$ best retained mobility. Representative high-mobility retained electron channels include Be$_3$C, AgBrO$_4$, Ta$_2$O$_5$ and TaF$_5$, whereas representative high-mobility retained hole channels include GeTe, GaN, SiSe and PbS (Fig.~\ref{fig:seed}c,d). The database supplies the two starting points for closed-loop design: QC-passed relaxed structures for generator adaptation and retained mobility entries for acquisition-ranker training. This split is important for the general framework. Structure feedback supplies the generator with examples that remain physically admissible after DFT relaxation, whereas channel feedback supplies the acquisition model with validated records worth prioritizing; neither stream alone is allowed to stand in for the other.

\subsection*{Retained evidence enriches budgeted DFT validation queues}

The retained seed evidence is useful if it can ration first-principles validation, even if it cannot replace it. This is the second test of the framework: whether retained feedback can guide what to calculate next while the authority to admit feedback remains with the first-principles gate. We therefore trained ALIGNN as an acquisition model whose purpose is to enrich a limited DFT validation queue while leaving final feedback assignment to first-principles validation. The target is the material-level $\log_{10}$ best retained mobility, defined as the largest retained channel mobility within each of the 280 QC-passed seed materials that also seed generator adaptation. A three-repeat, five-fold cross-validation grouped by reduced formula produces 840 out-of-fold predictions without formula-group leakage (Fig.~\ref{fig:screening}d--f and Supplementary Table~S5).

The ranker is not accurate enough to replace DFT, but it is informative enough to ration it. The out-of-fold rank correlations are Spearman $\rho=0.393$ and Kendall $\tau=0.269$, with MAE = 0.679 and RMSE = 0.920 in $\log_{10}$ mobility units. The absolute errors are too large for surrogate discovery, but the rank information prioritizes a small validation queue. The predicted top 10\% set recovers 8 of 28 true top-decile seed materials, a 2.9-fold enrichment over random selection ($P=2.9\times10^{-3}$, hypergeometric test; Fig.~\ref{fig:screening}f), while final feedback status remains assigned by first-principles validation.

The grouped validation design is important for interpreting this result. Materials with related compositions can otherwise appear in both training and test partitions, inflating apparent performance. By grouping folds by reduced formula, the evaluation asks whether the acquisition ranker can transfer ranking information across formula groups. The enrichment metric matches the use case more closely than a scalar regression score because a closed-loop campaign has a fixed DFT budget and must decide which few candidates should be validated next. In the framework logic, the ranker is therefore an allocator of validation effort rather than a source of new mobility labels.

\subsection*{Feedback-updated generation is compressed to a DFT-scale queue}

\begin{figure}[p]
\centering
\includegraphics[width=\textwidth]{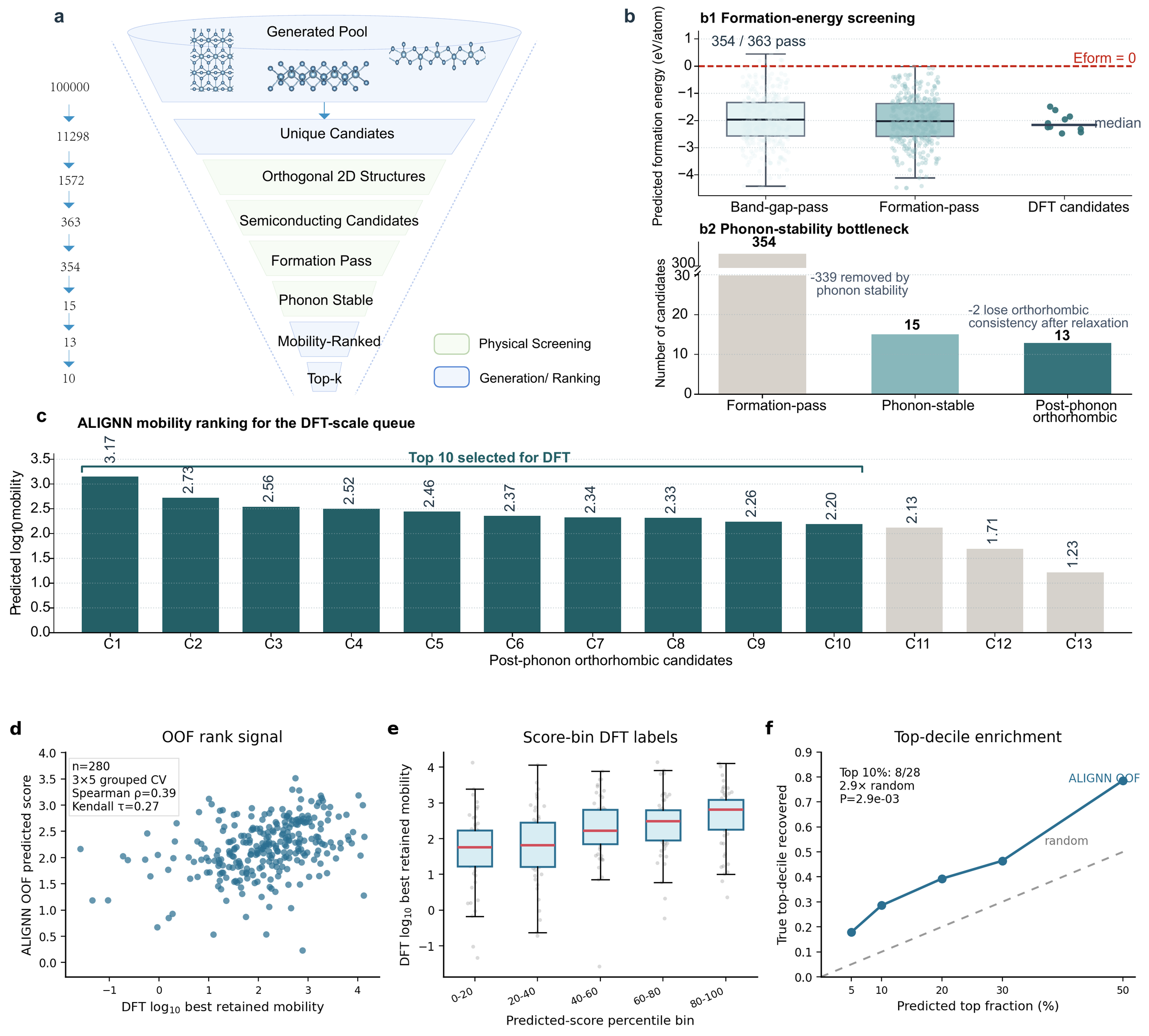}
\caption{\textbf{The framework compresses generated structures to DFT-scale validation queues without assigning feedback labels.} \textbf{a,} Screening funnel for a representative feedback-updated $10^5$-structure generation batch. \textbf{b,} Formation-energy and phonon-stability bottlenecks. Formation-energy screening produces a modest reduction, whereas dynamical-stability screening narrows the candidate set from 354 formation-pass candidates to 15 phonon-stable candidates. \textbf{c,} ALIGNN mobility ranking of the 13 post-phonon orthorhombic candidates; the top 10 are selected for DFT validation. \textbf{d,} Out-of-fold ALIGNN ranker scores from three-repeat, five-fold formula-grouped validation on 280 QC-passed seed materials. \textbf{e,} True DFT $\log_{10}$ best retained mobility distributions across predicted-score bins. \textbf{f,} True top-decile recovery among predicted top-ranked fractions. The dashed line denotes random expectation. ALIGNN is used for acquisition enrichment only; final feedback status is assigned by the first-principles validation engine.}
\label{fig:screening}
\end{figure}

Generative scale helps only after it is compressed into candidates worth calculating. To close the loop, feedback-updated structure generation is coupled to staged physical screening and acquisition ranking (Fig.~\ref{fig:screening}a). This stage answers a different question from the seed and acquisition analyses: not whether a mobility value is admissible, but whether a generated structure is sufficiently plausible to spend DFT mobility effort on it. The pipeline removes duplicate and reference-like structures, then applies orthorhombic/strict-90-degree structural filtering, semiconductor screening, formation-energy filtering, dynamical-stability screening, post-phonon orthorhombic checking and mobility-guided acquisition (Supplementary Note~5). Inexpensive structural, electronic and stability screens therefore define a plausible candidate set before the mobility ranker chooses a DFT-scale queue.

In a representative batch, this conversion is intentionally restrictive: 100,000 generated structures become 11,298 unique candidates, 1,572 orthorhombic candidates, 363 semiconducting candidates, 354 formation-pass candidates, 15 phonon-stable candidates and 13 post-phonon orthorhombic candidates. The largest reduction occurs at the dynamical-stability step: the formation-energy screen passes 354 of 363 semiconducting candidates, whereas phonon screening leaves only 15 stable candidates (Fig.~\ref{fig:screening}b). ALIGNN then ranks the 13 surviving candidates and selects the top 10 for first-principles validation (Fig.~\ref{fig:screening}c). Thus, the high-throughput part of the workflow creates a physically plausible DFT-scale queue; it does not assign the retained-feedback label.

This funnel clarifies the division of labour between generation and validation. The generator explores a broad structural space under feedback from retained seed and previous-loop structures, the physical filters remove invalid candidates, and the acquisition ranker operates only on candidates plausible enough to justify expensive transport validation. The design also prevents a common failure mode of closed-loop discovery: a model-generated structure is not treated as progress simply because it has a high predicted score. It becomes progress only if it survives physical screening and then produces reliability-gated first-principles feedback.

\subsection*{Iterative validation admits generated evidence back to the loop}

Closing the loop requires generated candidates to face the same evidence standard that created the seed data. This is where the framework differs from a one-way screen followed by retrospective reporting: validation results are sorted into feedback states before they are allowed to modify later rounds. Each shortlisted candidate was therefore returned to the first-principles mobility workflow used above. A generated formula enters the retained feedback set only if the workflow completes and at least one carrier--direction channel passes the final reliability gate, so each round of search can inform the next round without admitting prediction-only labels. A representative top-$k$ audit is shown in Supplementary Fig.~S3, with the validation-provenance definitions and full loop01--08 audit summarized in Supplementary Note~6, Supplementary Tables~S6 and S7 and Supplementary Data 4--15.

At campaign scale, the loop generated $2.4\times10^6$ structures and selected 102 candidates for first-principles validation under the same screening and acquisition protocol. The validation engine retained 41 generated follow-up formulas and 86 reliability-gated generated channels, expanding the retained mobility-feedback set from 573 seed channels to 659 total retained channels (Table~\ref{tab:campaign_accounting}; Fig.~\ref{fig:multiround}). Feedback accounting is chronological: only seed feedback and retained outputs from earlier completed iterations are available when a new acquisition state is built. Current-iteration DFT outcomes are admitted only after passing the validation gate, so the campaign remains a sequence of closed feedback episodes rather than a retrospective screen.

The low retained yield is an intended consequence of the evidence hierarchy. The generator and ranker enrich the validation queue, but the reliability gate decides which completed carrier--direction channels are allowed to change the next round. The retained set is therefore smaller than the calculated set by design, because the framework prioritizes feedback reliability over maximizing the number of reported finite mobilities.

\begin{table}[htbp]
\centering
\caption{\textbf{Framework accounting separates proposal, validation and retained feedback.} Counts distinguish the generated candidate pool, the first-principles validation queue and the retained feedback admitted only after channel-level reliability gating.}
\label{tab:campaign_accounting}
\small
\renewcommand{\arraystretch}{1.16}
\begin{tabularx}{\textwidth}{@{}>{\raggedright\arraybackslash}p{0.28\textwidth}>{\raggedright\arraybackslash}p{0.27\textwidth}>{\raggedright\arraybackslash}X@{}}
\toprule
Stage & Count & Role in feedback accounting \\
\midrule
Seed mobility construction & 516 materials; 2,064 attempted channels & Source pool for the first-principles mobility-feedback database. \\
Retained seed feedback & 280 QC-passed materials; 573 retained channels & Initial retained feedback used for acquisition-ranker training. \\
Generated campaign & $2.4\times10^6$ generated structures & Feedback-updated candidate pool screened before first-principles validation. \\
First-principles validation queue & 102 submitted candidates & DFT-scale queue selected by physical screening and acquisition ranking. \\
Retained generated feedback & 41 generated formulas; 86 reliability-gated channels & Generated feedback admitted after the final reliability gate. \\
Total retained feedback & 659 retained channels & Seed plus generated retained channels available after the campaign. \\
\bottomrule
\end{tabularx}
\end{table}

\subsection*{The validation scenario yields high-mobility retained channels}

\begin{figure}[htbp]
\centering
\includegraphics[width=\textwidth]{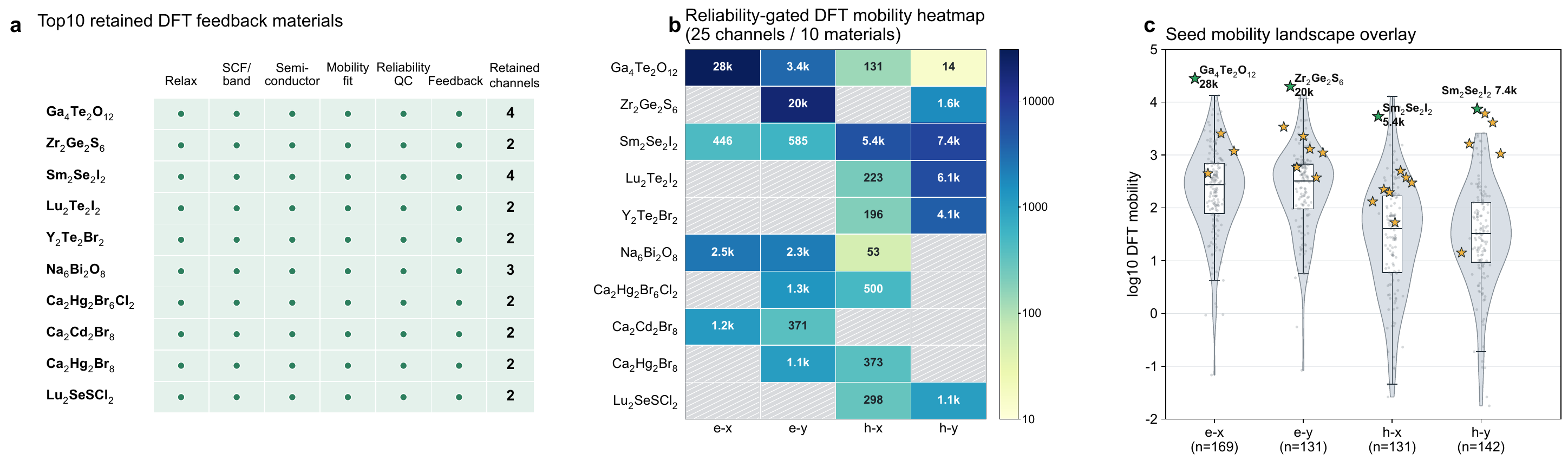}
\caption{\textbf{The mobility validation scenario yields high-mobility channels after reliability-gated first-principles validation.} \textbf{a,} Validation and feedback status of the top ten retained generated formulas. Dots denote completed workflow stages or passed checks, and the final column gives the number of retained carrier--direction channels. \textbf{b,} Retained DFT mobility heatmap for these top ten retained formulas. Hatched cells denote channels not retained as feedback. \textbf{c,} Retained generated channels overlaid on the retained seed-channel mobility distributions, emphasizing channel-level evidence rather than unconditional material rankings. Mobilities are reported in \cmvs.}
\label{fig:multiround}
\end{figure}

The generated examples show what the closed loop retained, not an unconditional ranking of materials. The retained generated set is intentionally small because prediction, physical screening and first-principles reliability gating must agree before feedback is admitted. Among the generated structures admitted by first-principles validation, the ten formulas with the highest retained channel mobilities provide a compact view of the retained generated evidence at the high-mobility end of the validation scenario (Fig.~\ref{fig:multiround}). Together, these ten formulas contain 25 retained carrier--direction channels and form the high-mobility subset of the 41 retained generated formulas and 86 reliability-gated generated channels reported above.

Figure~\ref{fig:multiround}a records the validation and feedback status of these ten generated formulas, and Fig.~\ref{fig:multiround}b reports their retained carrier--direction mobilities. The heatmap shows that retained feedback is channel-resolved: some formulas retain several electron and hole channels, whereas others contribute only one carrier type or one in-plane direction. Figure~\ref{fig:multiround}c compares these retained generated channels with the original retained seed-channel distributions.

The highest retained generated electron channels occur in Ga$_4$Te$_2$O$_{12}$ and Zr$_2$Ge$_2$S$_6$. Ga$_4$Te$_2$O$_{12}$ retains four reliability-gated channels, including an electron-$x$ mobility of approximately $2.8\times10^4$~\cmvs\ and an electron-$y$ mobility of approximately $3.4\times10^3$~\cmvs. Zr$_2$Ge$_2$S$_6$ contributes a high electron-$y$ mobility of approximately $2.0\times10^4$~\cmvs, together with a retained hole channel.

Sm$_2$Se$_2$I$_2$ has the highest retained generated hole-transport channels among the highlighted candidates. It retains all four carrier--direction channels, including hole mobilities of approximately $5.4\times10^3$ and $7.4\times10^3$~\cmvs. This result nominates Sm$_2$Se$_2$I$_2$ for follow-up hole-transport analysis, while leaving stronger material-level claims to additional electronic-structure checks.

Several highlighted candidates contain rare-earth elements, so their retained mobilities should be read as high-throughput feedback under the main validation protocol rather than final rare-earth transport predictions. The present mobility engine uses a uniform non-spin-polarized, no-DFT+$U$, no-SOC deformation-potential workflow to keep generated-channel feedback comparable across the campaign. For rare-earth compounds, localized $4f$ states, magnetic order and spin--orbit coupling may change the band-edge character and retained mobility channels. Band structures for the ten highlighted candidates, generated by the same multi-agent first-principles workflow, are provided in Supplementary Fig.~S5; the rare-earth follow-up rationale and a spin-polarized PBE+$U$ check for Sm$_2$Se$_2$I$_2$ are provided in Supplementary Note~8 and Supplementary Fig.~S4.

\section*{Discussion}

\textit{InvDesMobility} shifts closed-loop materials discovery from candidate generation alone to feedback admissibility. Its central object is a reusable feedback contract: an expensive first-principles value is not treated as a label until it has produced a reliability-resolved record that is allowed to teach the next round. This framing is useful beyond the specific mobility campaign because many target properties in computational materials science are assembled from multiple intermediate quantities and failure modes. A closed loop can therefore be technically active but scientifically brittle if it updates itself from every converged number. The framework addresses this problem at three levels: it coordinates first-principles execution at scale, records the intermediate evidence needed to judge a property value and admits only reliability-gated records back into the learning loop.

Carrier mobility in 2D semiconductors provides the validation case. In that campaign, 516 seed candidates yield a 573-channel mobility-feedback database across 280 QC-passed materials; retained evidence drives generation and acquisition; $2.4\times10^6$ generated structures are screened before 102 candidates are submitted to first-principles validation; and 41 generated formulas contribute 86 reliability-gated channels after revalidation. The main contribution is therefore a transferable framework for learning from expensive calculated properties, not a standalone ranking of candidates. The generated high-mobility examples are useful because they demonstrate that retained feedback can pass through the full loop, but their role is evidentiary rather than dispositive: they show what the framework admits under a defined high-throughput protocol, not final device-ready material claims.

The need for such a contract follows from the structure of the property itself. Mobility is assembled from band curvature, strain response, elastic stiffness, band-edge identity and fit reliability, so a finite value can still be a poor learning signal if any intermediate assumption fails. A generator can widen structural search and a ranker can enrich a validation queue, but neither decides whether a completed transport value is admissible evidence. The framework therefore places the strongest claim at the record level: a channel is useful feedback only after the physical workflow, provenance record and reliability gate agree. By making reliability a gating variable, \textit{InvDesMobility} connects 2D inverse design and generative materials discovery to provenance-aware and retrieval-supported scientific workflows\cite{Lookman2019ActiveLearning,DiffCSP2023,Merchant2023GNoME,Zeni2025MatterGen,Lyngby2022TwoDGenerative,Fung2021MatDesINNe,Jain2015FireWorks,Huber2020AiiDA,Hase2019SDL,MacLeod2020SDL,Boiko2023AutonomousChemicalResearch,Lewis2020RAG}.

The multi-agent layer is important because it coordinates evidence, not because it substitutes for physics. It keeps planning, execution, validation, recovery and provenance recording on the same stateful backbone, while deterministic first-principles outputs and reliability rules assign the final feedback status. This separation prevents surrogate scores from being treated as retained materials, prevents weak numerical outputs from entering the learning loop and keeps failed or withheld calculations in the audit trail. The agents expand the amount of auditable first-principles work that can be coordinated; the reliability gate determines which part of that work becomes scientific feedback.

Channel-level feedback is more faithful to 2D transport than a single material-level mobility number. One carrier direction may be reliable while the orthogonal direction is not; another material may have finite mobilities but only one direction supported by stable strain and effective-mass fits. Collapsing these cases to a best mobility would overstate what has been learned, whereas rejecting the whole material would discard usable evidence. The channel record is therefore the natural unit for mobility feedback, with material-level scores used only where a ranking task requires them.

This evidence design also changes how success should be read. A conventional screen often reports how many candidates exceeded a threshold after computation; here, the more important quantity is how many records survive the feedback gate and can therefore update later decisions. Retained, cautionary and withheld records are all part of the scientific output, but they play different roles. Retained records support learning, cautionary records expose fragile parts of the workflow, and withheld or failed records prevent the loop from mistaking completed computation for reliable evidence. This accounting is what makes the release auditable rather than merely reproducible from summary figures.

The same evidence standard also defines the study's limits. The workflow reports 300 K deformation-potential-limited mobilities from a high-throughput PBE protocol, not full electron--phonon Boltzmann mobilities\cite{Giustino2017ElectronPhonon,Ponce2020,Kaasbjerg2012MoS2}. The main calculation is non-spin-polarized and excludes DFT+$U$ and spin--orbit coupling; defects, substrates, contacts, grain boundaries, polar optical scattering and finite-temperature anharmonicity are also outside scope. These boundaries matter most for rare-earth candidates, where localized $4f$ states, magnetic order and spin--orbit coupling may alter band-edge character and retained mobility channels. The ungated-output comparison underlying this boundary is summarized in Supplementary Note~7 and Supplementary Table~S8. These limitations do not change the feedback logic, but they define the physical meaning of the retained labels produced in this validation scenario.

The closed-loop strategy has algorithmic limits as well. The generator is updated with QC-passed relaxed structures, while mobility values are routed to the acquisition ranker rather than used as direct generator targets. That choice reduces overfitting to a small mobility-feedback set, but leaves structural exploration and mobility optimization only indirectly coupled. The ranker is trained on 280 QC-passed seed materials, enough for enrichment but not broad extrapolation across all 2D chemistries, and strict orthorhombic filters exclude potentially interesting lower-symmetry semiconductors.

These limits point to a practical next stage. Narrowed candidate sets can be re-evaluated with electron--phonon scattering, spin--orbit coupling, DFT+$U$, magnetic order or defect-limited transport, while the loop itself can become uncertainty-aware and multi-objective across mobility, stability, synthesizability and novelty. The framework can also be extended to stronger property engines: replacing the deformation-potential workflow with more complete transport calculations would change the gate and the cost model, but not the requirement that only reliability-resolved records become feedback.

More broadly, feedback admissibility is a common bottleneck for expensive first-principles properties, including thermal conductivity, defect tolerance, excitonic response, catalytic activity and phonon stability, wherever the central question is not only what value was computed but whether that value should teach the next round. The contribution of \textit{InvDesMobility} is to make this question operational: it turns reliability from a post hoc discussion point into a record-level decision that controls learning. In that sense, the mobility campaign is a demanding validation scenario for a more general route to closed-loop materials discovery, where discovery progress is measured not only by proposed structures, but by the quality of the evidence admitted back into the loop.

\section*{Methods}

\textit{InvDesMobility} was implemented as a reusable closed-loop feedback framework and instantiated here as a carrier-mobility validation campaign. At the framework level, the same pattern can be reused for other expensive first-principles properties by replacing the property-specific intermediate quantities and gates while preserving proposal, validation, provenance and retained-feedback accounting. We therefore describe the method in two layers. The first layer is property-general: it defines how candidates are proposed, how calculations are launched, how provenance is retained, how outputs are stratified and how only retained records can update later rounds. The second layer is property-specific: it defines the mobility workflow, channel definitions and reliability gates used to test the framework in 2D semiconductors.

The learning unit was not a completed calculation, but a retained feedback record produced after deterministic mobility evaluation and reliability assessment. Relaxed structures and channel-resolved mobilities were routed as separate feedback objects: quality-controlled structures could update the generator, whereas mobility values could update acquisition only after channel-level reliability gating. This separation keeps structural admissibility, property reliability and validation priority from being collapsed into a single score. The protocol fixes the order of seed definition, runtime policy, first-principles mobility calculation, reliability assessment, generation and acquisition, acquisition-ranker validation and round-wise feedback accounting. A mobility value entered the learning set only after passing the reliability rules below.

Each feedback decision is reproducible from stored records. Material identity, structure source, calculation status, fitted physical quantities, uncertainty estimates, quality labels, acquisition scores and round-wise feedback state were recorded as separate fields. Figures and tables were derived from these records rather than manually edited candidate lists. The evidence hierarchy separates acquisition scores for validation priority, deterministic DFT outputs for calculated physical quantities, cautionary or withheld records for audit and reliability-gated retained channels for learning feedback; this separation is the operational definition of feedback admissibility used throughout the campaign. In practical terms, a record can be high-scoring, completed or numerically finite without being retained feedback; retention requires the additional evidence checks described below.

\subsection*{Seed dataset and channel definitions}
Seed structures were selected from 2DMatPedia using a 2D semiconductor pre-screening procedure\cite{Zhou2019TwoDMatPedia}. The database provides a bottom-up and top-down reference pool for 2D materials and is complementary to other high-throughput 2D-materials resources\cite{Mounet2018Exfoliation,Haastrup2018C2DB}. After structural and electronic filtering, 516 candidates were passed to the mobility feedback engine. For each material, four in-plane transport channels were defined: electron-$x$, electron-$y$, hole-$x$ and hole-$y$. The $x$ and $y$ directions were assigned from standardized in-plane lattice vectors and used consistently for seed and generated structures.

The channel was the basic unit of mobility feedback throughout the study. This choice follows from the anisotropic nature of 2D transport: electron and hole channels, and the two in-plane directions, can fail or pass reliability checks independently. A material-level target was used only when training or evaluating the acquisition ranker; in that case, the target was the best retained channel mobility for a material after applying the reliability gate. Materials without at least one retained channel were excluded from the acquisition-ranker training target but remained in the audit records. Full seed records, channel definitions and material-level labels are described in Supplementary Note~1 and provided in Supplementary Data 1--3.

\subsection*{Runtime architecture and control policy}
The multi-agent mobility framework uses a typed shared state to link workflow planning, tool execution, validation, recovery and feedback assignment to the same VASP-based mobility backbone. The shared state is the mechanism that turns automation into evidence accounting: the same material record carries planning decisions, executed tool calls, recovery attempts, intermediate files, final fitted quantities and the feedback status assigned after validation. The orchestration layer combines a typed state graph, persistent checkpointing, structured tool outputs, recovery and validation records, VASP-native tool calls and a common result schema. In batch operation, candidates were claimed from a validation queue, VASP inputs were prepared, the same single-material runner was invoked, and completed, failed or skipped outcomes were recorded in the shared schema. This design follows the same motivation as provenance-aware workflow systems: every output should be traceable to a concrete computational state and execution path\cite{Jain2015FireWorks,Huber2020AiiDA}.

The retrieval-supported policy layer provides stage-scoped context for VASP-writing stages, including relaxation, SCF, band-structure and strain-loop substages. Within these stages, it can propose INCAR or KPOINTS adjustments and recovery actions\cite{Lewis2020RAG}. Retrieval draws on a curated VASP policy corpus and a cleaned VASP Wiki corpus; the runtime falls back to fixed templates if retrieved evidence or model output is inadequate. Skill packages support planning, critique, recovery, validation, reporting and execution-feasibility checks, but final fitted quantities and retained-channel status are assigned downstream by the deterministic mobility workflow and reliability rules. Thus, the agentic layer increases the amount of coordinated and recoverable first-principles work, whereas the physics workflow and reliability gates determine what becomes scientific feedback.

\subsection*{First-principles mobility workflow}
All main-workflow first-principles mobility calculations used VASP with the projector augmented-wave method and the PBE exchange-correlation functional\cite{Kresse1996VASP,Kresse1999PAW,Perdew1996PBE}. Calculations were non-spin-polarized and did not include spin--orbit coupling unless explicitly stated for follow-up analysis. PBE PAW potentials were used. 2D structures were treated as slab systems with one $k$ point along the out-of-plane direction. Unless otherwise stated, ENCUT was 600 eV, ISMEAR = 0, SIGMA = 0.01 eV, ALGO = Normal, NELM = 300, NELMIN = 4 and IVDW = 12\cite{Grimme2010D3,Grimme2011D3BJ}. Relaxations used EDIFF = $10^{-5}$ eV, EDIFFG = $-0.01$ eV \AA$^{-1}$, IBRION = 2, POTIM = 0.5 and NSW = 200. SCF, band and effective-mass stages used EDIFF = $10^{-6}$ eV. Band structures were evaluated along $\Gamma$--X--S--Y--$\Gamma$ with 40 points per segment. A compact settings table and deterministic workflow schematic are provided in Supplementary Table~S2 and Supplementary Fig.~S1.

Stage-specific KPOINTS policies and slab settings were handled by the workflow configuration and recorded in the calculation provenance, so the reported mobility entries can be traced back to their exact VASP input state rather than to manually edited labels.

The workflow order was fixed for all main calculations. Each candidate was first relaxed, then evaluated by static self-consistent and band-structure calculations. Effective masses were extracted near the relevant band edges, after which uniaxial strain calculations were performed along the two in-plane directions. Deformation-potential slopes and 2D elastic constants were obtained from these strain series. Mobility was assembled only after all required intermediate quantities were available for the corresponding channel.

Carrier mobility was evaluated using the 2D deformation-potential-limited expression\cite{Bardeen1950,Ponce2020,Zhang2023Mobility}
\begin{equation}
\mu_{c,\alpha}=\frac{e\hbar^3 C_{\alpha}^{2D}}{k_{\mathrm B}T\,m^*_{c,\alpha}m_{c,d}E_{1,c,\alpha}^{2}},
\end{equation}
where $c$ denotes carrier type, $\alpha$ denotes transport direction, $C_{\alpha}^{2D}$ is the 2D elastic constant, $m^*_{c,\alpha}$ is the transport effective mass, $m_{c,d}$ is the density-of-states effective mass and $E_{1,c,\alpha}$ is the deformation-potential constant. Mobilities were evaluated at 300 K. Effective masses were fitted from 21 points with 0.01 \AA$^{-1}$ spacing within $|k-k_0|\leq0.03$ \AA$^{-1}$. Deformation-potential and elastic fits used uniaxial strains from $-0.020$ to 0.020 in steps of 0.005, with band-edge energies aligned to the vacuum level before fitting. The deformation-potential model provides a computationally tractable high-throughput property record for the validation scenario, while more complete electron--phonon approaches remain the appropriate follow-up for narrowed candidate sets\cite{Giustino2017ElectronPhonon,Ponce2020,Kaasbjerg2012MoS2}.

\subsection*{Reliability assessment}
Reliability was assigned at the carrier--direction channel level before any record was used for training or acquisition feedback. The gate was designed to test the evidence supporting a mobility value rather than the magnitude of the value itself. Seed channels were retained if they had completed workflow status, positive finite mobility, finite $E_1$ and $C^{2D}$, at least seven completed strain points, $\min(R^2_{E_1},R^2_{C^{2D}})\geq0.985$, relative uncertainty of $E_1\leq0.25$, relative uncertainty of $C^{2D}\leq0.10$, $|E_1|\geq0.10$ eV, effective-mass fit $R^2\geq0.95$ and effective mass between $0.02m_0$ and $20m_0$. Caution-level seed channels used weaker thresholds, including at least five strain points, $\min(R^2_{E_1},R^2_{C^{2D}})\geq0.95$, relative uncertainty of $E_1\leq0.50$, relative uncertainty of $C^{2D}\leq0.20$ and $|E_1|\geq0.05$ eV.

Generated channels were retained under the final feedback rule only when the workflow completed, mobility was positive and finite, key fitted quantities were finite, fitting checks passed, effective mass was accepted and no dynamic band-edge switch was detected. Where checks corresponded to the same fitted quantities, generated-channel criteria followed the same physical logic as the seed workflow; generated-candidate-specific thresholds and exclusions are listed in Supplementary Tables~S3 and S4. These criteria define whether a numerical output becomes feedback.

The reliability statuses were used asymmetrically. Retained channels entered the retained mobility-feedback set and could be used for acquisition-ranker training. Caution-level and weak channels remained available for audit, threshold-sensitivity analysis and failure-mode inspection but were not used as retained feedback. Failed or non-usable records documented workflow attrition, allowing a calculation to be scientifically informative without being admissible as a training signal. This asymmetric use is essential for closed-loop learning: failed or fragile calculations should inform the audit trail and future protocol design, but they should not update the acquisition model as if they were reliable property labels. Material-level QC classes were derived from channel statuses: a material was QC-passed if at least one channel was retained, caution-level if no channel was retained but at least one channel reached the caution tier, and not retained otherwise.

\subsection*{Generation, screening and acquisition}
Structure generation used DiffCSP, following the InvDesFlow implementation\cite{DiffCSP2023,InvDesFlow2024,InvDesFlowAL2025}. The model was fine-tuned with QC-passed relaxed structures during feedback updates, while mobility values were not used as direct generator targets. This choice keeps the generator focused on producing structurally admissible candidates and leaves mobility optimization to screening, acquisition and first-principles validation. Generator updates used only feedback available before the corresponding generation round; current-round validation outcomes were excluded until the round closed. Generated structures were then passed through a fixed physical-screening pipeline before mobility acquisition ranking: duplicate removal, orthorhombic/strict-90-degree filtering, ALIGNN electronic screening, MEGNet formation-energy screening, PhononBench dynamical-stability screening, post-phonon orthorhombic consistency checking and ALIGNN mobility acquisition ranking\cite{Choudhary2021ALIGNN,Chen2019MEGNet,Togo2015Phonopy,PhononBench2025}. Duplicate removal used lattice tolerance 0.2, site tolerance 0.3 and angle tolerance 5.0$^\circ$ with pymatgen\cite{Ong2013Pymatgen}. The formation-energy threshold was $<0.0$ eV atom$^{-1}$, and the phonon-stability workflow used a $2\times2\times1$ supercell. The ranker was applied only after these filters, ensuring that acquisition scores selected among plausible validation candidates rather than replacing the physical screens.

A generation batch can sample 100,000 candidate structures. The candidate list was filtered stepwise by applying the screens in a fixed order, and only post-phonon orthorhombic candidates were passed to the mobility acquisition ranker. The fixed ordering prevents candidates with high predicted mobility scores that fail basic physical checks from entering the validation queue. Top-$k$ candidates were then submitted to the first-principles mobility feedback engine, where final retained feedback entries were assigned independently of the acquisition score.

\subsection*{ALIGNN acquisition-ranker validation}
The ALIGNN mobility model was evaluated as an acquisition ranker, not as a final mobility predictor (Fig.~\ref{fig:screening}d--f). The validation used the 280 QC-passed seed materials with the material-level target $\log_{10}$ best retained mobility. We performed three repeats of five-fold grouped cross-validation and grouped materials by reduced formula to avoid splitting related formula groups across train, validation and test partitions. Predictions were recorded only for held-out test materials and averaged over the three out-of-fold predictions per material. Performance was assessed using Spearman rank correlation, Kendall rank correlation, MAE, RMSE and top-decile enrichment. For held-out material $i$, let $y_i$ denote the DFT $\log_{10}$ best retained mobility and $\hat{y}_i$ the out-of-fold ALIGNN prediction. The pointwise errors were summarized as
\begin{equation}
\mathrm{MAE}=\frac{1}{n}\sum_i |\hat{y}_i-y_i|,\qquad
\mathrm{RMSE}=\left[\frac{1}{n}\sum_i(\hat{y}_i-y_i)^2\right]^{1/2}.
\end{equation}
Spearman $\rho$ was computed from the predicted and true ranks, whereas Kendall $\tau$ measured pairwise rank concordance. Top-decile enrichment reports the fraction of true top-decile seed materials recovered within predicted top-ranked fractions.

The top-decile enrichment calculation treated the 28 highest-mobility seed materials as the true top decile. For a predicted top fraction $f$, we counted how many of these 28 materials appeared within the highest-ranked $f$ fraction of predictions and compared the result with random expectation. The hypergeometric test for the predicted top 10\% set used the full QC-passed seed population as the background population. This evaluation was chosen because the acquisition problem is a budgeted ranking problem: the model is useful if it enriches a small validation queue for candidates likely to yield retained high-mobility evidence, even when pointwise mobility errors remain large.

\subsection*{Round-wise feedback accounting}
Feedback was accounted round by round to prevent current-loop validation results from leaking into current-loop acquisition. This accounting rule makes the campaign a sequence of prospective feedback episodes rather than a retrospective relabeling of a completed screen. For loop $N$, before-round feedback counts are the seed retained channels and structures plus retained generated feedback from loops earlier than $N$. Current-loop validation results are counted only after the loop completes and are not available to the generator update or acquisition ranking that selected the same loop's candidates. Runtime manifest counts are reported separately as audit metadata and are not used to define retained feedback.

Two counts were therefore maintained for each round. The first count describes the validation queue and includes submitted, completed, failed and skipped first-principles jobs. The second count describes feedback admitted to the learning loop and includes only retained formulas and retained carrier--direction channels. Keeping these counts separate prevents completed but non-retained numerical outputs from being mistaken for successful feedback and ensures that the cumulative label count corresponds to the actual training signal available after each round; the same rule is used in the main figures, Supplementary Tables~S6 and S7 and the released feedback records.

\section*{Data availability}
All source data, retained feedback records and workflows are released at \url{https://github.com/DreamLufei/invDesMobility}, with a companion evidence website at \url{https://dreamlufei.github.io/invDesMobility/} for interactive exploration of screening funnels, validation provenance and reliability-gated mobility feedback. Processed source data are also archived on Zenodo at \url{https://doi.org/10.5281/zenodo.20475023}. The release is organized as an audit trail for the feedback contract rather than only as input for reproducing summary figures. It links seed material records, channel-level mobility records with retained, cautionary and withheld states, ALIGNN validation inputs and predictions, round-wise feedback accounting, top-$k$ validation provenance, literature comparison records, a file manifest and SHA-256 checksums in CSV, JSON and SQLite formats; these files are described in Supplementary Table~S10. This organization makes the transition from candidate, to screening state, to validation status, to retained feedback decision inspectable at the record level. VASP and licensed pseudopotential files are proprietary and are not redistributed; large raw VASP working directories, OUTCAR, CHGCAR, WAVECAR, licensed POTCAR files and local credentials are not redistributed.

\section*{Code availability}
Workflow code is available in three public GitHub repositories: \texttt{invDesMobility} for generation, screening and acquisition ranking (\url{https://github.com/DreamLufei/invDesMobility}), \texttt{2d-mobility} for the multi-agent first-principles mobility runtime (\url{https://github.com/DreamLufei/2d-mobility}) and \texttt{invdesmobility\_loop} for feedback extraction and closed-loop orchestration (\url{https://github.com/DreamLufei/invdesmobility_loop}). The public main-branch snapshots checked for this submission are \texttt{f8aeec6}, \texttt{48f01dc} and \texttt{80f17f6}, respectively. The repositories include source code, environment files, step-level scripts, dry-run checks and reproduction notes, and are released under the MIT License.

\section*{Acknowledgements}
The work is supported by the National Natural Science Foundation of China (Nos. 62476278 and 11934020), Beijing Natural Science Foundation (No. Z250005) and the National Key R\&D Program of China (Grant No. 2024YFA1408601). Computational resources were provided by the Physical Laboratory of High Performance Computing at Renmin University of China.

\section*{Author contributions}
W.-K.L. and Z.-F.G. conceived the project. W.-K.L. implemented the workflow, performed first-principles calculations and analysed the data. Z.-F.G., W.J. and P.-J.G. contributed to workflow design, materials screening and interpretation. Z.-F.G. and Z.-Y.L. supervised the project. All authors contributed to manuscript preparation.

\section*{Competing interests}
The authors declare no competing interests.

\section*{AI-assisted writing}
External AI-assisted writing tools, separate from the reported workflow, were used only for language editing and manuscript organization under author supervision.

\section*{Supplementary information}
Supplementary materials are available for this paper. Supplementary Notes~1--8 provide seed-record definitions, runtime details, screening and validation provenance, rare-earth follow-up checks and boundary analyses. Supplementary Tables~S1--S10, Supplementary Figs.~S1--S5 and Supplementary Data~1--33 support the workflow description, mobility-feedback records, screening pipeline, reliability criteria, validation accounting and data manifest.

\bibliographystyle{unsrtnat}
\bibliography{references}

\clearpage
\setcounter{section}{0}
\setcounter{figure}{0}
\setcounter{table}{0}
\setcounter{equation}{0}

\renewcommand{\thesection}{S\arabic{section}}
\renewcommand{\thefigure}{S\arabic{figure}}
\renewcommand{\thetable}{S\arabic{table}}
\renewcommand{\theequation}{S\arabic{equation}}

\input{SI_arxiv}

\end{document}

%% file: SI_arxiv.tex
\begin{center}
{\Large \textbf{Supplementary Information}}\\[0.5em]
{\large A reliability-gated first-principles feedback framework for closed-loop materials discovery}\\[0.5em]
Wen-Kao Li, Ze-Feng Gao, Peng-Jie Guo, Wei Ji and Zhong-Yi Lu
\end{center}

\tableofcontents
\clearpage

\section*{Supplementary Methods and Notes}
\addcontentsline{toc}{section}{Supplementary Methods and Notes}

\subsection*{Supplementary Note 1: Seed records and channel definitions}
The seed set contains 516 2DMatPedia-derived two-dimensional semiconductor candidates after structural and electronic pre-screening. Each material was assigned four in-plane carrier--direction channels: electron-$x$, electron-$y$, hole-$x$ and hole-$y$. The complete material-level seed summaries are provided in Supplementary Data 1, and the channel-level table with all 2,064 attempted channels is provided in Supplementary Data 2. The retained seed-channel landscape used for Fig. 2 and for generated-channel comparison is provided in Supplementary Data 3.

A material-level quality tier is computed from channel-level reliability labels. A material is QC-passed when it contains at least one retained channel, caution-level when it contains no retained channel but at least one caution-level channel, and not retained when none of the four channels reaches either level. This rule yields 280 QC-passed materials, 83 caution-level records and 153 not-retained records. The target used for acquisition-ranker training and validation is the material-level $\log_{10}$ best retained mobility, defined as the maximum retained channel mobility within a QC-passed material followed by a base-10 logarithm.

\subsection*{Supplementary Note 2: Implementation modules and reproducibility boundaries}
The implementation is organized into three functional code modules that support the reliability-gated feedback framework in Fig. 1. The first module is the stateful first-principles mobility runtime, which executes single-material and batch mobility calculations, stores typed workflow state, tracks calculation artifacts and records validation outcomes. The second module is the generation, screening and acquisition pipeline, which builds mobility seed and feedback datasets, adapts the DiffCSP/InvDesFlow generator, generates candidate structures, deduplicates candidates, applies physical screens, ranks survivors with ALIGNN and exports strict-90-degree top-$k$ validation queues. The third module is the campaign bridge, which extracts retained feedback from completed mobility batches, archives QC-passed relaxed structures, builds feedback-aware DiffCSP and ALIGNN datasets and launches the next validation round.

This modular split is used for reproducibility and audit, not as a scientific claim in itself. Large generated pools, trained checkpoints, raw VASP directories and licensed VASP inputs are not redistributed. The processed data, figure source tables, round-wise accounting and SQLite databases needed to verify the manuscript figures and tables are supplied through the Zenodo and GitHub source-data records described in the Data availability and Code availability statements, not through the manuscript TeX-source bundle.

\subsection*{Supplementary Note 3: Agentic policy, RAG and skill packages}
The agentic policy layer provides stage-scoped support for VASP-writing stages: relaxation, SCF, band-structure and VASP-writing substages inside the strain loop. It can propose selected stage-scoped INCAR overrides, KPOINTS policy choices, failure-diagnosis evidence and recovery suggestions.

The RAG layer retrieves from two corpora: a bundled house-policy corpus distilled from project defaults and recovery priors, and a cleaned VASP Wiki corpus stored in Postgres and indexed with pgvector. When the policy layer is active, trace artifacts such as \path{retrieval_trace.json}, \path{parameter_plan.json} and \path{recovery_diagnosis.json} can be emitted. Evidence-poor or malformed decisions fall back to fixed templates. Disk-backed skill packages provide summary-first, on-demand context for single-material mobility calculation, batch mobility screening, recovery, strain refinement, physics validation, reporting, admission, planning, critique, orchestration, cost guarding and execution feasibility. Skill use is recorded in \path{skill_trace} artifacts. Supplementary Fig.~S2 illustrates the context-window behaviour: compact skill hints appear first, and detailed SKILL.md content is loaded only when a stage-specific decision requires it.

\subsection*{Supplementary Note 4: Deterministic mobility chain}
Each material-level calculation follows an ordered state sequence: structure preparation, relaxation, self-consistent electronic-structure calculation, band-structure calculation, semiconductor verification, effective-mass sampling, strain calculations, deformation-potential fitting, elastic fitting, mobility evaluation, channel-level reliability assessment and feedback assignment. Planner, critic, orchestrator, execution, validation and recovery agents operate on this state, but numerical quantities are generated only by deterministic first-principles calculations and post-processing. Supplementary Fig.~S1 summarizes the deterministic dataset and mobility-calculation path.

\subsection*{Supplementary Note 5: Generation and physical screening pipeline}
The generation/screening route is: DiffCSP/InvDesFlow generation; generated-set self-deduplication and deduplication against the 516 seed/reference structures; orthorhombic or strict-90-degree filtering; ALIGNN electronic screening; MEGNet formation-energy prediction and filtering at $E_{\mathrm{form}}<0$ eV atom$^{-1}$; PhononBench dynamical-stability screening; post-phonon strict-90-degree validation; ALIGNN mobility ranking; and top-$k$ CIF export. The PhononBench workflow performs relaxation, phonopy displacement-supercell construction, force-constant assembly and phonon-spectrum analysis. The phonon-stability gate is the dominant physical filtering step in the feedback-updated generation round.

\subsection*{Supplementary Note 6: Generated-proposal validation provenance}
The representative top-$k$ validation map in Supplementary Fig.~S3 is retained as a provenance example because it shows the separation between ALIGNN acquisition and first-principles feedback assignment. The updated source-data package reports the full loop01--08 generated-proposal audit after relaxed-structure deduplication: 106 raw VASP workflow rows are reduced to 102 representative submitted candidates, yielding 41 retained generated formulas and 86 reliability-gated generated carrier--direction channels. Candidate-level validation outcomes, retained channel-resolved mobility values, rejected-feedback records and duplicate-structure clusters are provided in Supplementary Data 4--15. The multi-round accounting tables in Supplementary Data 25--27 then connect these validation records to the chronological feedback state used by the loop.

\subsection*{Supplementary Note 7: Reliability-gating and literature sanity checks}
The current data do not provide a same-candidate pure-script versus agentic-runtime ablation. Instead, the available retrospective control is a reliability-gating audit: how many finite or otherwise usable mobility outputs would have entered feedback without channel-level quality control. Supplementary Table~S8 reports this ungated-output comparison. Literature sanity-check records are provided in Supplementary Table~S9 and Supplementary Data 30. These records were not used for training, acquisition ranking or quality assignment.

\subsection*{Supplementary Note 8: Rare-earth candidate follow-up}
The main high-throughput validation workflow uses a non-spin-polarized, no-SOC PBE deformation-potential protocol. Rare-earth candidates retained by this protocol require higher-level follow-up because localized $4f$ states and spin--orbit coupling may alter band-edge character and transport. Sm$_2$Se$_2$I$_2$ is therefore treated as a retained rare-earth candidate under the main validation protocol rather than as a definitive final material claim. Supplementary Fig.~S4 shows a spin-polarized PBE+$U$ band-structure check as an example follow-up analysis, and Supplementary Fig.~S5 reports the main-workflow band structures for the ten highlighted generated formulas.

\section*{Supplementary Tables}
\addcontentsline{toc}{section}{Supplementary Tables}

\begin{table}[H]
\centering
\caption{Functional implementation modules and framework relevance. The public code repositories are described in the Code availability statement.}

\label{tab:module_map}
\small
\begin{tabularx}{\textwidth}{@{}p{0.24\textwidth}X X@{}}
\toprule
Module & Main role & Manuscript relevance \\
\midrule
Mobility runtime & Stateful LangGraph-based first-principles carrier-mobility runtime; typed shared state; VASP-native tools; checkpoint/store support; RAG policy layer; disk-backed skills; human-in-the-loop resume. & Implements the multi-agent orchestration layer, VASP-based mobility backbone and runtime roles in Fig. 1b. \\
Generation, screening and acquisition & Dataset preparation, DiffCSP/InvDesFlow generator adaptation, $10^5$-scale generation, deduplication, orthorhombic filtering, ALIGNN electronic screening, MEGNet formation screening, PhononBench stability screening, ALIGNN ranking and top-$k$ export. & Implements the generation, physical-screening and acquisition path in Fig. 3. \\
Closed-loop bridge & Extracts retained feedback from completed mobility batches; builds feedback-aware DiffCSP and ALIGNN datasets; publishes top-$k$ validation queues; launches downstream validation. & Implements the round-wise feedback accounting summarized in Fig. 4 and Table 2. \\
\bottomrule
\end{tabularx}
\end{table}

\begin{table}[H]
\centering
\caption{DFT and mobility-workflow settings used in the main validation protocol.}
\label{tab:dft_settings}
\small
\begin{tabularx}{\textwidth}{@{}p{0.33\textwidth}X@{}}
\toprule
Setting & Value \\
\midrule
Code & VASP \\
Exchange--correlation functional & PBE \\
PAW potentials & PBE PAW potentials; POTCAR files not redistributed \\
Spin treatment & Non-spin-polarized in the main workflow \\
Spin--orbit coupling & Not included in the main high-throughput workflow \\
Plane-wave cutoff & 600 eV \\
Smearing & ISMEAR = 0, SIGMA = 0.01 eV \\
Electronic minimization & ALGO = Normal, NELM = 300, NELMIN = 4 \\
Dispersion correction & IVDW = 12 \\
Relaxation convergence & EDIFF = $10^{-5}$ eV; EDIFFG = $-0.01$ eV \AA$^{-1}$ \\
Relaxation algorithm & IBRION = 2, POTIM = 0.5, NSW = 200 \\
SCF/band/effective-mass convergence & EDIFF = $10^{-6}$ eV \\
Band path & $\Gamma$--X--S--Y--$\Gamma$ \\
Line-mode sampling & 40 points per segment \\
Out-of-plane sampling & $k_z=1$ \\
Mobility temperature & 300 K \\
\bottomrule
\end{tabularx}
\end{table}

\begin{table}[H]
\centering
\caption{Seed-channel reliability thresholds. These criteria define feedback admissibility rather than mobility magnitude.}
\label{tab:seed_qc}
\small
\begin{tabularx}{\textwidth}{@{}p{0.42\textwidth}XX@{}}
\toprule
Criterion & Retained channel & Caution-level channel \\
\midrule
Completed strain points & $\geq 7$ & $\geq 5$ \\
$\min(R^2_{E_1},R^2_{C^{2D}})$ & $\geq0.985$ & $\geq0.95$ \\
Relative uncertainty of $E_1$ & $\leq0.25$ & $\leq0.50$ \\
Relative uncertainty of $C^{2D}$ & $\leq0.10$ & $\leq0.20$ \\
$|E_1|$ & $\geq0.10$ eV & $\geq0.05$ eV \\
Effective-mass fit & $R^2\geq0.95$ & $R^2\geq0.95$ \\
Effective-mass range & $0.02m_0$--$20m_0$ & $0.02m_0$--$20m_0$ \\
\bottomrule
\end{tabularx}
\end{table}

\begin{table}[H]
\centering
\caption{Generated-channel retained-feedback thresholds. A generated channel was retained as reliability-gated mobility feedback only if the workflow completed and these checks passed.}
\label{tab:generated_qc}
\small
\begin{tabularx}{\textwidth}{@{}p{0.42\textwidth}X@{}}
\toprule
Criterion & Threshold or requirement \\
\midrule
Mobility & Positive and finite \\
$E_1$ and $C^{2D}$ & Finite \\
$\min(R^2_{E_1},R^2_{C^{2D}})$ & $\geq0.90$ \\
Relative uncertainty of $E_1$ & $\leq0.50$ \\
Relative uncertainty of $C^{2D}$ & $\leq0.20$ \\
$|E_1|$ & $\geq0.05$ eV \\
Effective-mass fit & $R^2\geq0.95$ \\
Effective-mass range & $0.02m_0$--$20m_0$ \\
Maximum branch energy jump & $\leq0.50$ eV \\
Dynamic band-edge switch & Not detected \\
\bottomrule
\end{tabularx}
\end{table}

\begin{table}[H]
\centering
\caption{ALIGNN acquisition-ranker validation metrics. The model was evaluated as a ranker, not as a final mobility predictor.}
\label{tab:alignn_metrics}
\small
\begin{tabular}{@{}ll@{}}
\toprule
Metric & Value \\
\midrule
Materials & 280 QC-passed seed materials \\
Grouping & Reduced formula (\path{formula_reduced_abc}) \\
Cross-validation & 3 repeats $\times$ 5 folds \\
Out-of-fold predictions & 840, averaged by material \\
Spearman $\rho$ & 0.393 \\
Kendall $\tau$ & 0.269 \\
MAE in $\log_{10}$ mobility & 0.679 \\
RMSE in $\log_{10}$ mobility & 0.920 \\
Predicted top 10\% recovery & 8/28 true top-decile materials \\
Top-10\% enrichment & 2.9$\times$ random; $P=2.9\times10^{-3}$ \\
\bottomrule
\end{tabular}
\end{table}

\begin{table}[H]
\centering
\caption{Full round-wise inverse-design funnel and retained first-principles feedback accounting. Current-loop validation outputs are reported as after-round outcomes and are not included in the same loop's before-round feedback counts. Runtime manifest counts are audit metadata only.}
\label{tab:round_full}
\scriptsize
\setlength{\tabcolsep}{3.0pt}
\textbf{a, Generation and screening denominators}\par\vspace{0.4em}
\resizebox{\textwidth}{!}{%
\begin{tabular}{@{}lrrrrrrrrr@{}}
\toprule
Round & Labels before & Struct. before & Generated & Unique & Ortho. & Semi. & Form. & Phonon & Post-phonon ortho. \\
\midrule
Seed & 573 & 280 & -- & -- & -- & -- & -- & -- & -- \\
Loop 1 & 573 & 280 & 100{,}000 & 11{,}298 & 1{,}572 & 363 & 354 & 15 & 13 \\
Loop 2 & 582 & 284 & 100{,}000 & 12{,}494 & 2{,}095 & 446 & 438 & 14 & 10 \\
Loop 3 & 585 & 285 & 100{,}000 & 12{,}818 & 2{,}100 & 460 & 451 & 8 & 6 \\
Loop 4 & 588 & 286 & 100{,}000 & 17{,}164 & 3{,}250 & 727 & 705 & 15 & 11 \\
Loop 5 & 597 & 289 & 500{,}000 & 15{,}352 & 2{,}612 & 1{,}576 & 1{,}533 & 27 & 20 \\
Loop 6 & 623 & 300 & 500{,}000 & 21{,}962 & 5{,}222 & 3{,}322 & 3{,}243 & 73 & 53 \\
Loop 7 & 631 & 304 & 500{,}000 & 15{,}581 & 4{,}002 & 2{,}528 & 2{,}456 & 62 & 49 \\
Loop 8 & 636 & 307 & 500{,}000 & 22{,}944 & 5{,}393 & 3{,}478 & 3{,}387 & 84 & 63 \\
\bottomrule
\end{tabular}%
}
\vspace{1.0em}
\textbf{b, DFT validation and feedback accounting}\par\vspace{0.4em}
\resizebox{\textwidth}{!}{%
\begin{tabular}{@{}lrrrrrrrrrrrr@{}}
\toprule
Round & Sub. uniq. & Sub. raw & Completed & Dup. rm. & Retained F. & Retained ch. & Best $\mu_e$ & Best $\mu_h$ & Channels added & Struct. added & Cum. ch. & Cum. struct. \\
\midrule
Seed & -- & -- & -- & -- & -- & -- & -- & -- & 0 & 0 & 573 & 280 \\
Loop 1 & 10 & 10 & 10 & 0 & 4 & 9 & 211.4 & 723.3 & 9 & 4 & 582 & 284 \\
Loop 2 & 9 & 10 & 8 & 1 & 1 & 3 & 2536.2 & 52.6 & 3 & 1 & 585 & 285 \\
Loop 3 & 3 & 6 & 3 & 3 & 1 & 3 & 641.9 & 470.8 & 3 & 1 & 588 & 286 \\
Loop 4 & 10 & 10 & 10 & 0 & 3 & 9 & 584.8 & 7406.2 & 9 & 3 & 597 & 289 \\
Loop 5 & 20 & 20 & 18 & 0 & 11 & 26 & 1101.5 & 908.2 & 26 & 11 & 623 & 300 \\
Loop 6 & 10 & 10 & 10 & 0 & 4 & 8 & 28027.3 & 1002.2 & 8 & 4 & 631 & 304 \\
Loop 7 & 10 & 10 & 9 & 0 & 3 & 5 & 10342.4 & 397.1 & 5 & 3 & 636 & 307 \\
Loop 8 & 30 & 30 & 30 & 0 & 14 & 23 & 19754.4 & 6054.6 & 23 & 14 & 659 & 321 \\
\bottomrule
\end{tabular}%
}
\begin{flushleft}\footnotesize Ortho. denotes the orthorhombic or strict-90-degree structural filter; Semi. denotes the semiconductor screen; Form. denotes the formation-energy screen. Sub. uniq. gives the relaxed-structure-deduplicated DFT validation queue; Sub. raw gives raw VASP workflow rows before duplicate removal. Best mobilities are retained channel maxima in cm$^2$ V$^{-1}$ s$^{-1}$.\end{flushleft}
\end{table}

\begin{table}[H]
\centering
\caption{Runtime validation audit summary for the loop01--08 top-$k$ validation queues after relaxed-structure deduplication. Completed but not retained means that a DFT validation job completed but did not return a formula-level retained feedback record.}
\label{tab:runtime_audit}
\small
\resizebox{\textwidth}{!}{%
\begin{tabular}{@{}lrrrrrrrr@{}}
\toprule
Round & Sub. uniq. & Sub. raw & DFT completed & Failed/skipped & Dup. rm. & Retained formulas & Retained channels & Completed but not retained \\
\midrule
Loop 1 & 10 & 10 & 10 & 0 & 0 & 4 & 9 & 6 \\
Loop 2 & 9 & 10 & 8 & 1 & 1 & 1 & 3 & 7 \\
Loop 3 & 3 & 6 & 3 & 0 & 3 & 1 & 3 & 2 \\
Loop 4 & 10 & 10 & 10 & 0 & 0 & 3 & 9 & 7 \\
Loop 5 & 20 & 20 & 18 & 2 & 0 & 11 & 26 & 7 \\
Loop 6 & 10 & 10 & 10 & 0 & 0 & 4 & 8 & 6 \\
Loop 7 & 10 & 10 & 9 & 1 & 0 & 3 & 5 & 6 \\
Loop 8 & 30 & 30 & 30 & 0 & 0 & 14 & 23 & 16 \\
\midrule
Total & 102 & 106 & 98 & 4 & 4 & 41 & 86 & 57 \\
\bottomrule
\end{tabular}%
}
\end{table}

\begin{table}[H]
\centering
\caption{Retrospective reliability-gating audit. Usable channels are finite or otherwise analyzable numerical outputs before the strict retained-feedback gate. Retained channels are the subset admitted as reliability-gated mobility feedback.}
\label{tab:reliability_gate_audit}
\small
\begin{tabular}{@{}lrrrrr@{}}
\toprule
Dataset & Usable channels & Retained & Withheld usable & Caution & Weak \\
\midrule
Seed mobility-feedback database & 1{,}894 & 573 & 1{,}321 & 301 & 1{,}020 \\
Generated-proposal audit & 390 & 86 & 304 & 4 & 300 \\
\bottomrule
\end{tabular}
\begin{flushleft}
\footnotesize Failed or non-usable channels are excluded from the usable-channel denominator but are retained in the audit database: 170 seed channels and 18 generated-proposal channels. The generated-proposal row uses the loop01--08 deduplicated validation audit. This table supports the claim that numerical completion is not equivalent to retained feedback.
\end{flushleft}
\end{table}

\begin{table}[H]
\centering
\caption{Literature-anchored sanity checks for retained seed channels. These comparisons were not used for training, acquisition ranking or quality assignment. Mobilities are in cm$^2$ V$^{-1}$ s$^{-1}$; ratio denotes this work divided by the reported literature value.}
\label{tab:literature_sanity}
\scriptsize
\setlength{\tabcolsep}{3.2pt}
\resizebox{\textwidth}{!}{%
\begin{tabular}{@{}lccrrrll@{}}
\toprule
Material & Carrier & Direction & This work & Literature & Ratio & Phase consistency & Confidence \\
\midrule
PbS & hole & $y$ & 2601.2 & 2190 & 1.19 & strong & high \\
GeSe & hole & $x$ & 155.7 & 119 & 1.31 & strong & high \\
SnSe & electron & $x$ & 1807.2 & 2390 & 0.76 & plausible & medium-high \\
SnTe & hole & $y$ & 1854.7 & 3440 & 0.54 & plausible & medium \\
GeTe & electron & $y$ & 3357.3 & 1590 & 2.11 & plausible & medium \\
SnS & electron & $y$ & 766.3 & 1930 & 0.40 & plausible & medium-low \\
GeSe & electron & $x$ & 7095.3 & 6220 & 1.14 & strong & medium-high (caution channel) \\
GeSe & electron & $y$ & 12238.3 & 29300 & 0.42 & strong & medium (caution channel) \\
\bottomrule
\end{tabular}%
}
\begin{flushleft}
\footnotesize PbS was compared with a few-layer PbX first-principles report; GeSe and SnSe-family checks use reported $\beta$-phase or group-IV monochalcogenide literature values. Experimental field-effect mobilities and intrinsic deformation-potential-limited mobilities are not directly equivalent, so these records are used only as sanity checks of scale and phase family.
\end{flushleft}
\end{table}

{\scriptsize\sloppy
\begin{longtable}{@{}p{0.06\textwidth}p{0.38\textwidth}p{0.50\textwidth}@{}}
\caption{Supplementary Data files supplied through the cleaned Zenodo/GitHub source-data package. These records are listed for reproducibility and are not required for compiling the manuscript TeX-source bundle.}\label{tab:data_list}\\
\toprule
No. & File & Purpose \\
\midrule
\endfirsthead
\toprule
No. & File & Purpose \\
\midrule
\endhead
1 & \path{data/seed_materials_summary.csv} & Seed material summaries, material-level reliability tiers and training targets. \\
2 & \path{data/seed_mobility_channels.csv} & Full seed channel-level mobility records and channel QC labels. \\
3 & \path{data/seed_retained_channel_landscape.csv} & Retained seed-channel landscape used for Fig. 2 and Fig. 4 overlays. \\
4 & \path{data/topk_validation_status.csv} & Loop01--08 deduplicated validation status table for 102 submitted generated proposals. \\
5 & \path{data/topk_trusted_channels.csv} & Deduplicated reliability-gated generated-channel records; 86 retained channels enter feedback. \\
6 & \path{data/topk_all_channel_review.csv} & All 408 deduplicated generated-channel review records, including rejected and warning-level outputs. \\
7 & \path{data/topk_trusted_materials.csv} & Retained generated-material summaries for 41 generated formulas. \\
8 & \path{data/topk_rejected_feedback.csv} & Non-retained generated proposals after deduplicated validation. \\
9 & \path{data/topk_feedback_summary.json} & Loop01--08 feedback-assignment summary and deduplication counts. \\
10 & \path{data/topk_all_channel_review_raw106.csv} & Raw pre-deduplication channel review records for 106 VASP workflow rows. \\
11 & \path{data/topk_trusted_channels_raw87.csv} & Raw pre-deduplication retained-channel records. \\
12 & \path{data/loop01_08_all_106_mobility_review_table.csv} & Raw material-level VASP review table before relaxed-structure deduplication. \\
13 & \path{data/loop01_08_dedup_102_mobility_review_table.csv} & Material-level VASP review table after relaxed-structure deduplication. \\
14 & \path{data/loop01_08_duplicate_structure_clusters.csv} & Four duplicate relaxed-structure rows removed from the raw validation table. \\
15 & \path{data/loop01_08_all_106_mobility_simple_table.csv} & Compact raw material-level validation table for inspection. \\
16 & \path{data/alignn_seed_targets.csv} & ALIGNN ranker targets for 280 QC-passed seed materials. \\
17 & \path{data/alignn_id_prop.csv} & Two-column ALIGNN/JARVIS id--property input table. \\
18 & \path{data/alignn_cv_splits.csv} & 3$\times$5 formula-grouped CV train/validation/test assignments. \\
19 & \path{data/alignn_cv_predictions.csv} & 840 out-of-fold ALIGNN predictions. \\
20 & \path{data/alignn_cv_predictions_aggregated.csv} & 280 material-level averaged ALIGNN predictions and prediction-score bins. \\
21 & \path{data/alignn_validation_metrics.json} & Overall ALIGNN ranker metrics. \\
22 & \path{data/alignn_predicted_quantile_bins.csv} & True mobility statistics by predicted-score bin. \\
23 & \path{data/alignn_top_decile_enrichment.csv} & Top-decile enrichment source data. \\
24 & \path{data/alignn_cv_audit.json} & Formula-group leakage audit for the repeated grouped CV. \\
25 & \path{data/multi_round_feedback_accounting_main.csv} & Main-text round-wise feedback accounting table for eight feedback iterations. \\
26 & \path{data/multi_round_feedback_accounting_full.csv} & Full loop01--08 round-wise feedback accounting table. \\
27 & \path{data/multi_round_feedback_sources.json} & Field definitions and source manifests for round-wise accounting. \\
28 & \path{data/literature_mobility_matches_2026-03-31.csv} & Original literature sanity-check records. \\ 
29 & \path{data/reliability_gate_audit.csv} & Retrospective ungated-output versus retained-feedback audit. \\ 
30 & \path{data/literature_sanity_checks.csv} & Cleaned literature sanity-check table used in Supplementary Table~S9. \\ 
31 & \path{data/threshold_sensitivity_analysis.csv} & Threshold-sensitivity reclassification records. \\ 
32 & \path{data/ncs_mobility_submission_used.sqlite} & Combined SQLite database containing seed, ALIGNN, literature and updated loop01--08 generated-proposal tables. \\ 
33 & \path{data/ncs_loop01_08_vasp_dedup.sqlite} & Lightweight SQLite copy of the updated loop01--08 VASP validation tables. \\ 
\bottomrule
\end{longtable}
}

\clearpage
\section*{Supplementary Figures}
\addcontentsline{toc}{section}{Supplementary Figures}

\begin{figure}[H]
\centering
\includegraphics[width=0.82\textwidth]{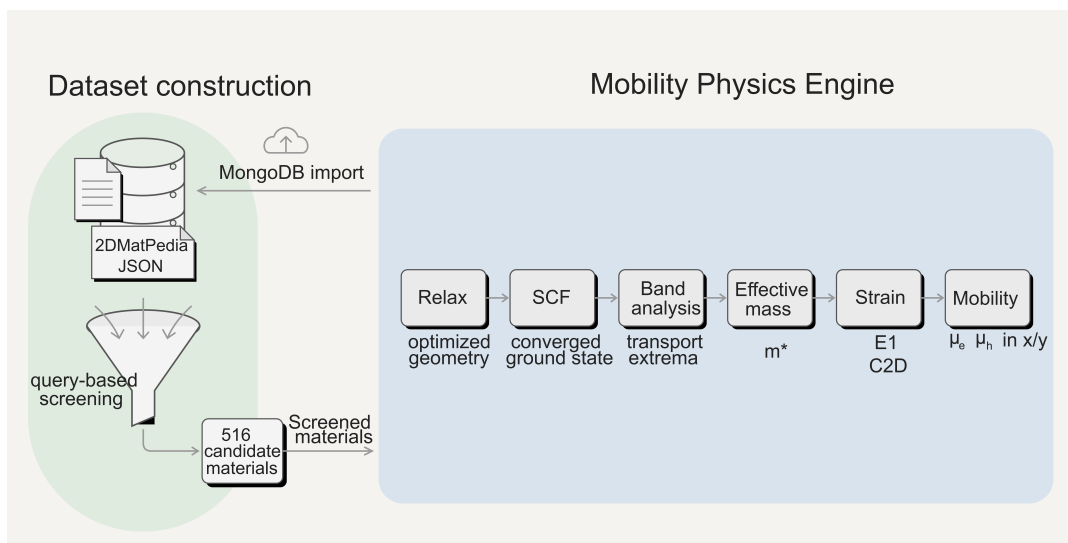}
\caption{\textbf{Seed dataset construction and deterministic mobility-workflow schematic.} The workflow connects 2DMatPedia import, structure standardization, first-principles mobility calculation, channel-level reliability assessment and material-level aggregation.}
\label{fig:S1}
\end{figure}

\begin{figure}[H]
\centering
\includegraphics[width=0.98\textwidth]{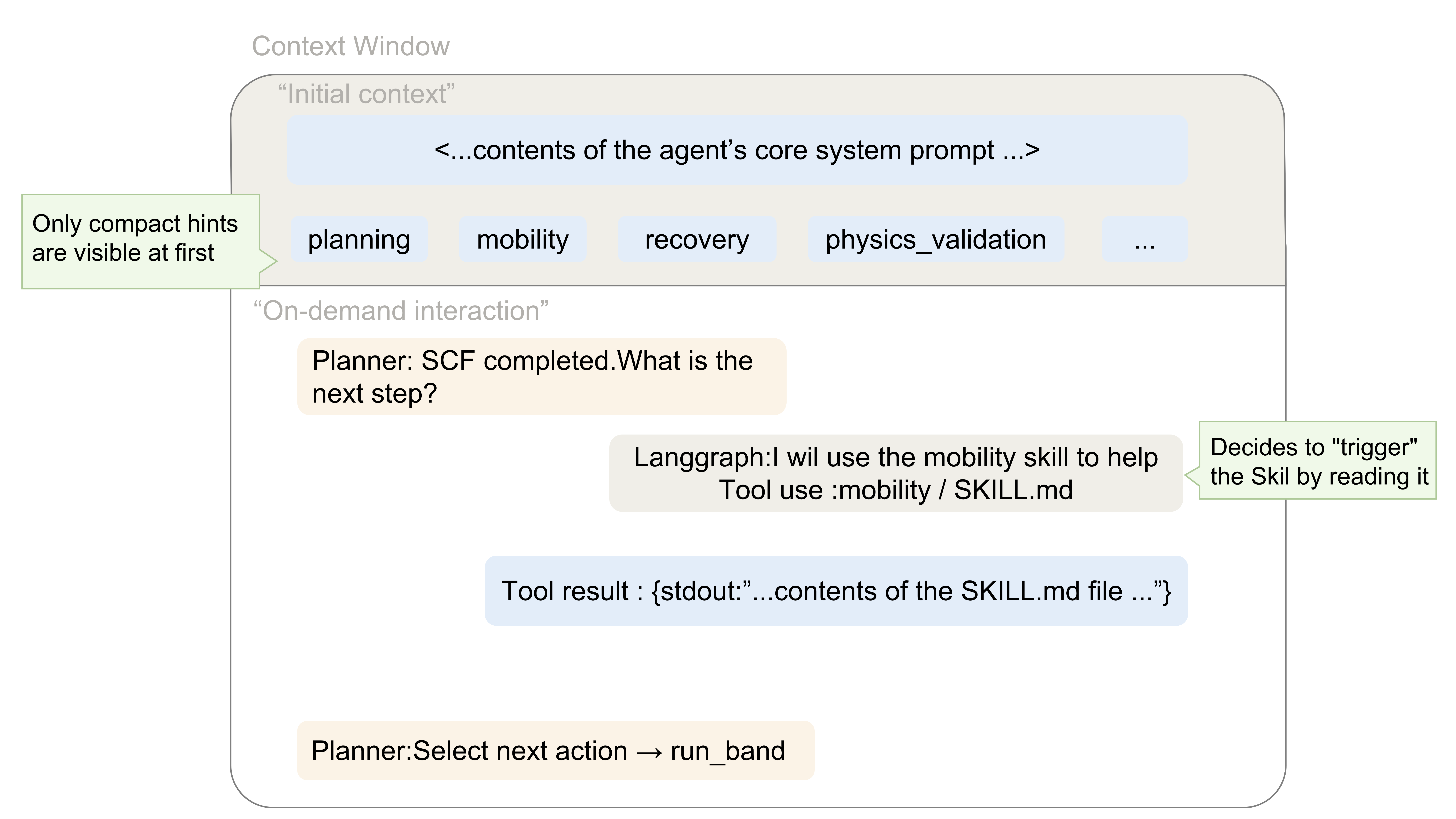}
\caption{\textbf{On-demand skill-package loading in the agentic runtime.} Compact skill hints are available in the initial agent context, while detailed stage-specific SKILL.md content is loaded only when the current state requires it. This keeps detailed policy context available for planning, validation, recovery and reporting without loading every skill package into every stage.}
\label{fig:S2}
\end{figure}

\begin{figure}[H]
\centering
\includegraphics[width=0.98\textwidth]{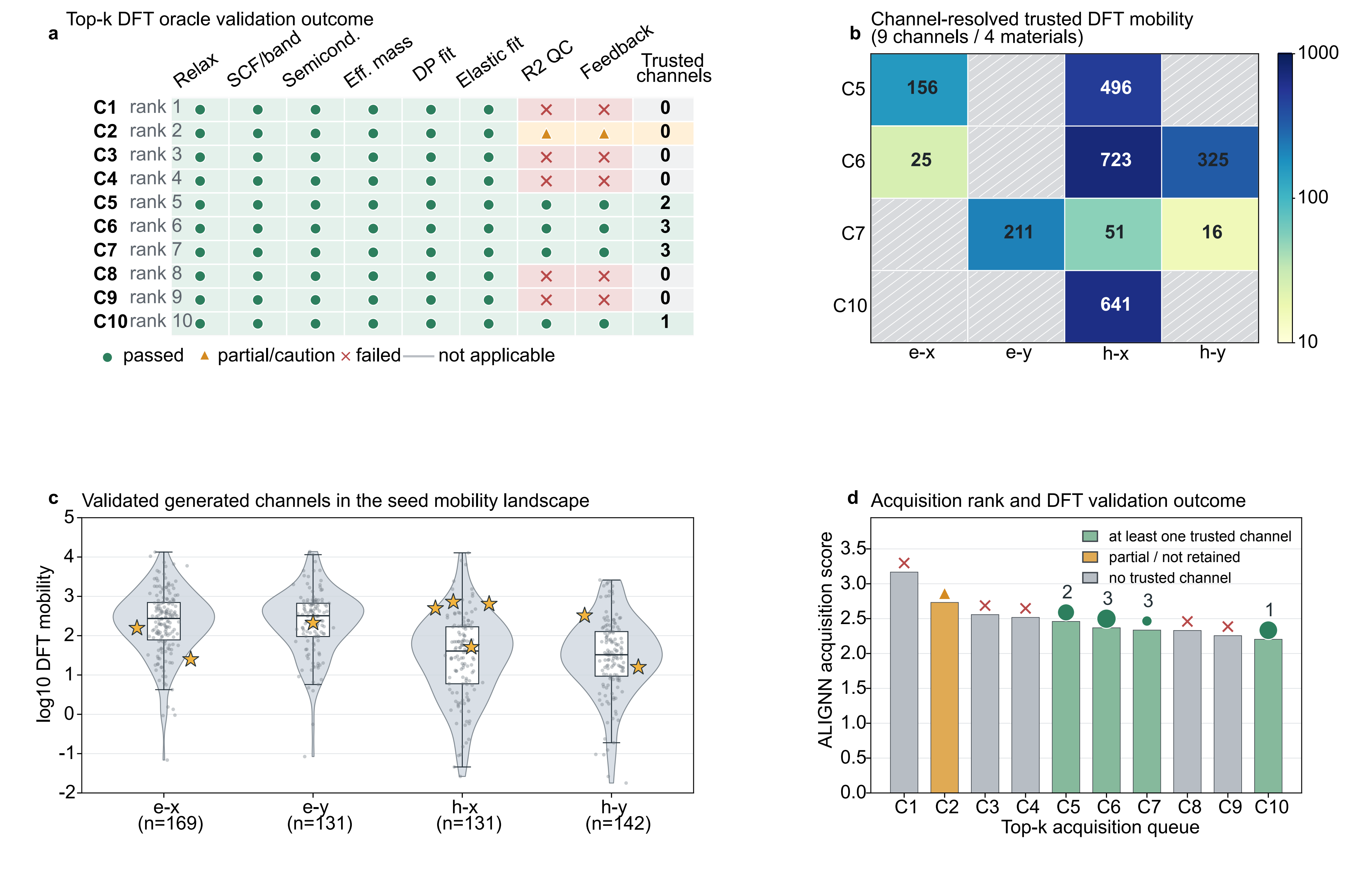}
\caption{\textbf{Representative top-$k$ validation provenance.} Candidate-level validation outcomes, retained channel-resolved mobility values, seed-database overlay and ALIGNN acquisition scores illustrate how acquisition ranking, DFT workflow status and channel-level QC are kept separate. The updated source-data tables provide the full loop01--08 validation audit.}
\label{fig:S3}
\end{figure}

\begin{figure}[H]
\centering
\includegraphics[width=0.75\textwidth]{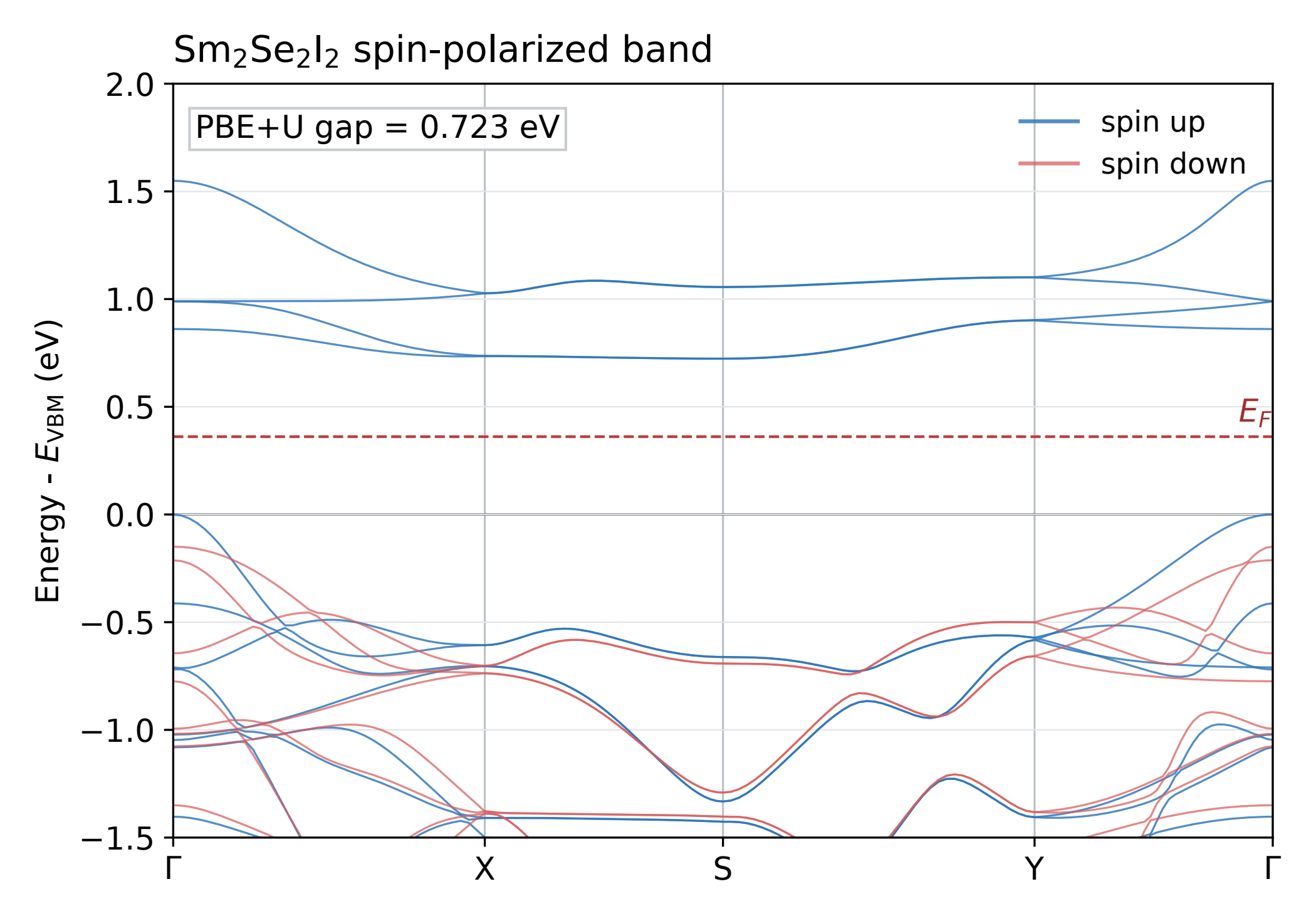}
\caption{\textbf{Spin-polarized PBE+$U$ band-structure check for Sm$_2$Se$_2$I$_2$.} The rare-earth candidate is retained by the main no-SOC validation protocol but requires localized-$f$ and spin--orbit robustness checks before stronger material-level claims are made.}
\label{fig:S4}
\end{figure}

\begin{figure}[H]
\centering
\setlength{\tabcolsep}{2pt}
\renewcommand{\arraystretch}{0.92}
\begin{tabular}{ccc}
\textbf{a} Ga$_4$Te$_2$O$_{12}$ & \textbf{b} Zr$_2$Ge$_2$S$_6$ & \textbf{c} Sm$_2$Se$_2$I$_2$ \\
\includegraphics[width=0.30\textwidth]{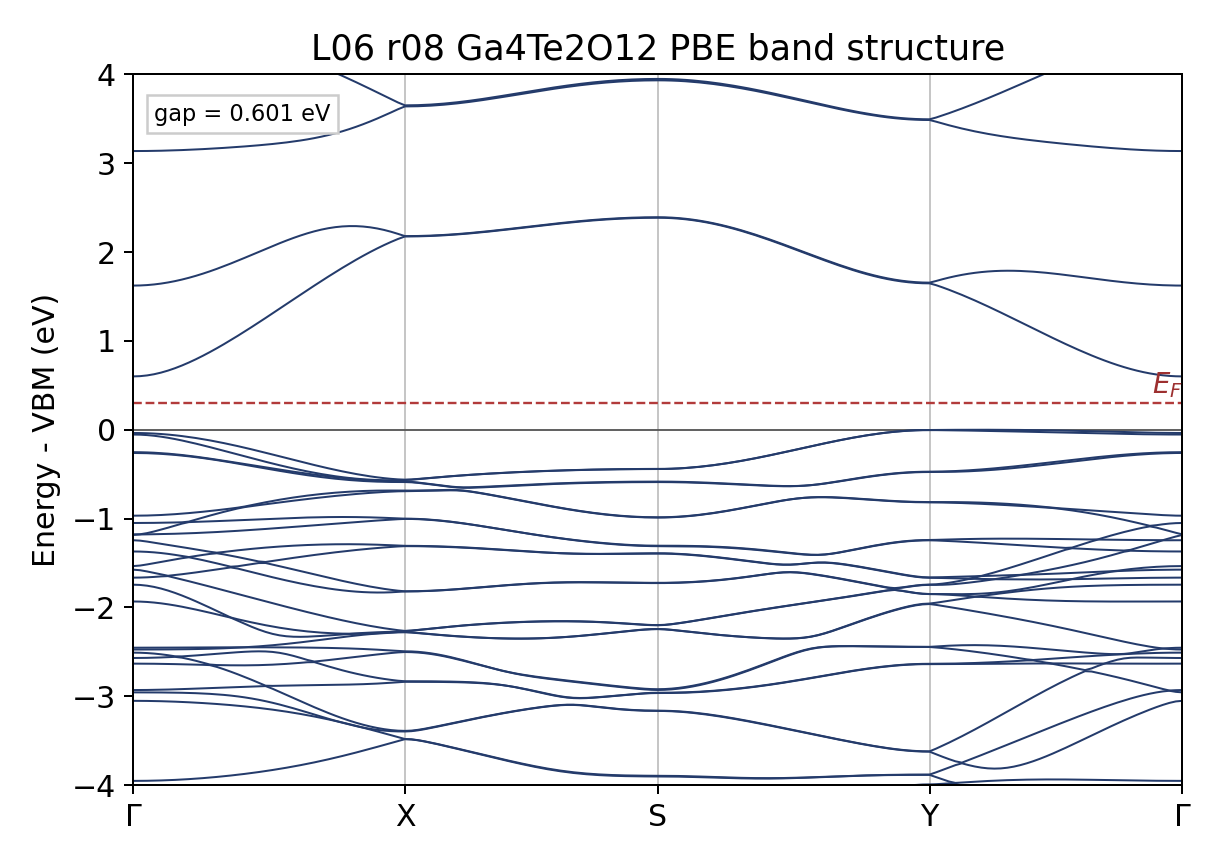} &
\includegraphics[width=0.30\textwidth]{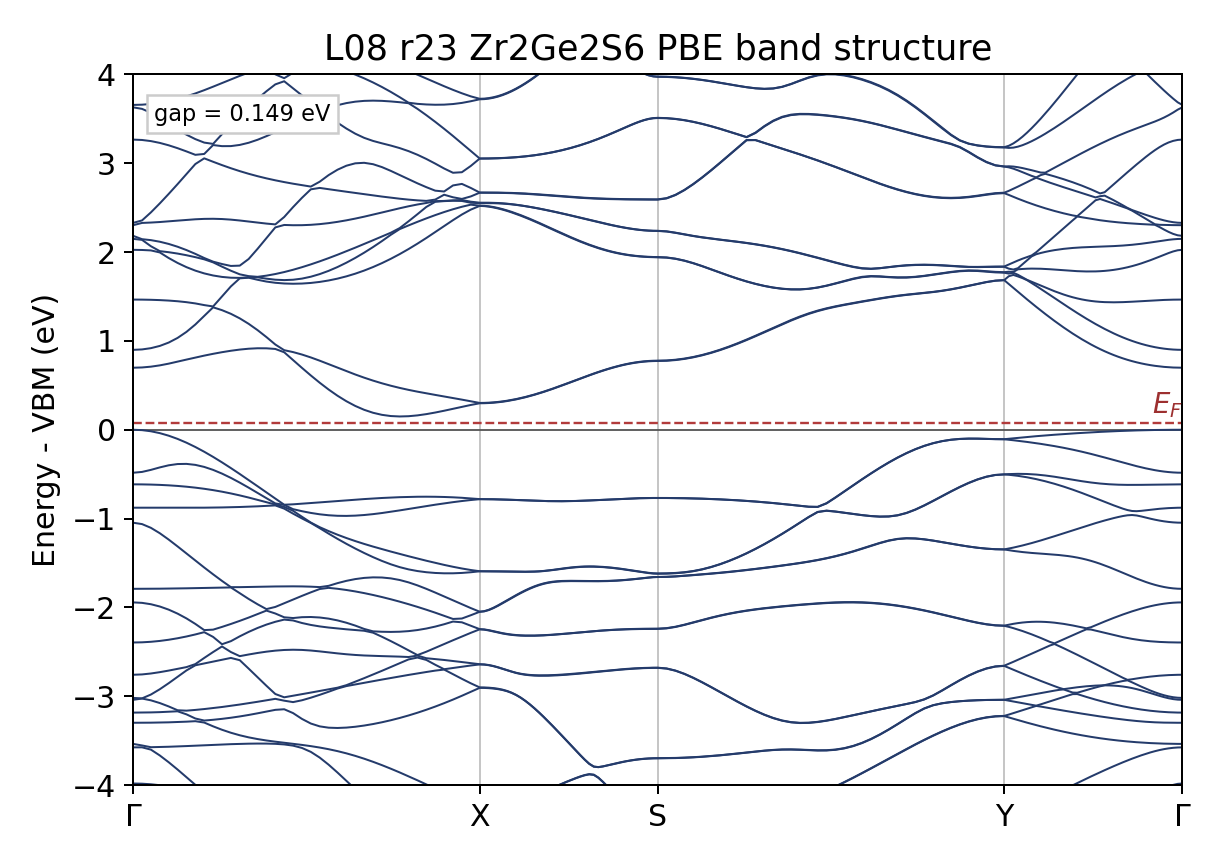} &
\includegraphics[width=0.30\textwidth]{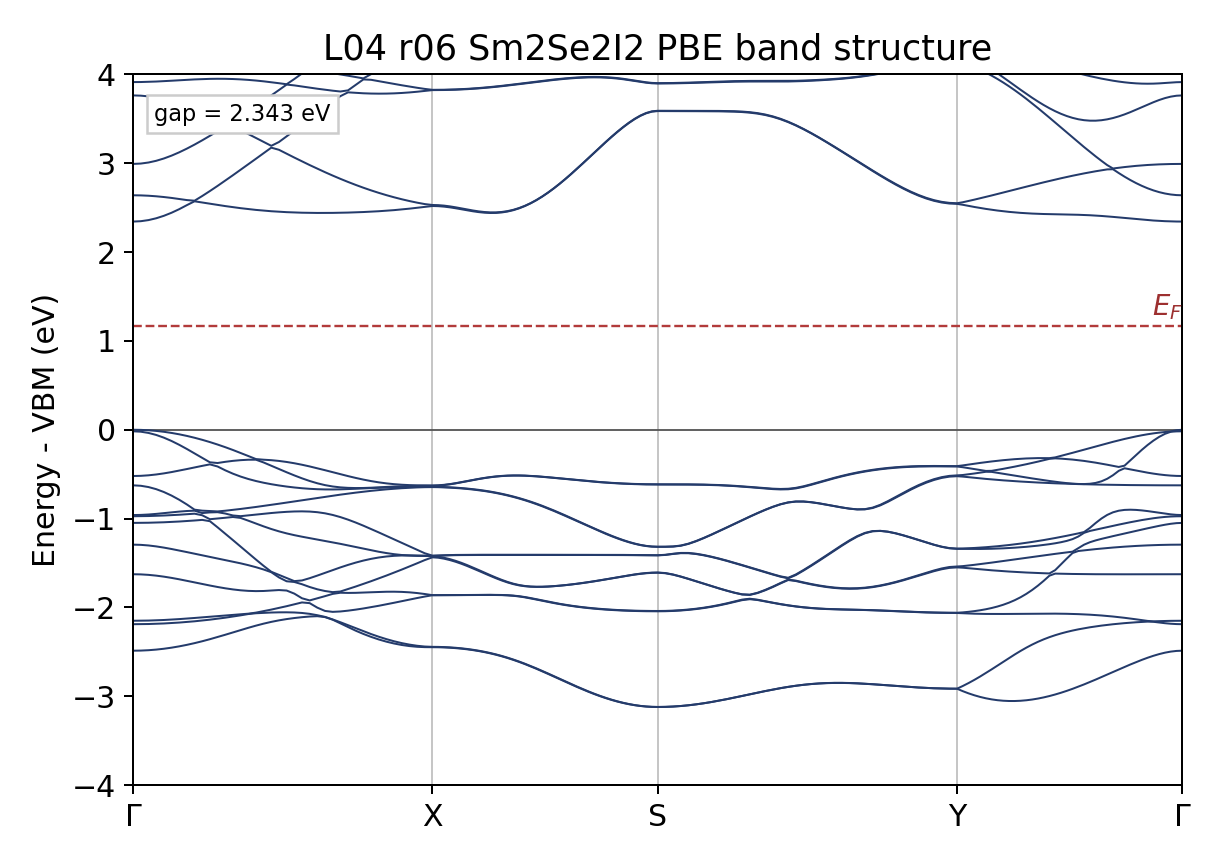} \\
\textbf{d} Lu$_2$Te$_2$I$_2$ & \textbf{e} Y$_2$Te$_2$Br$_2$ & \textbf{f} Na$_6$Bi$_2$O$_8$ \\
\includegraphics[width=0.30\textwidth]{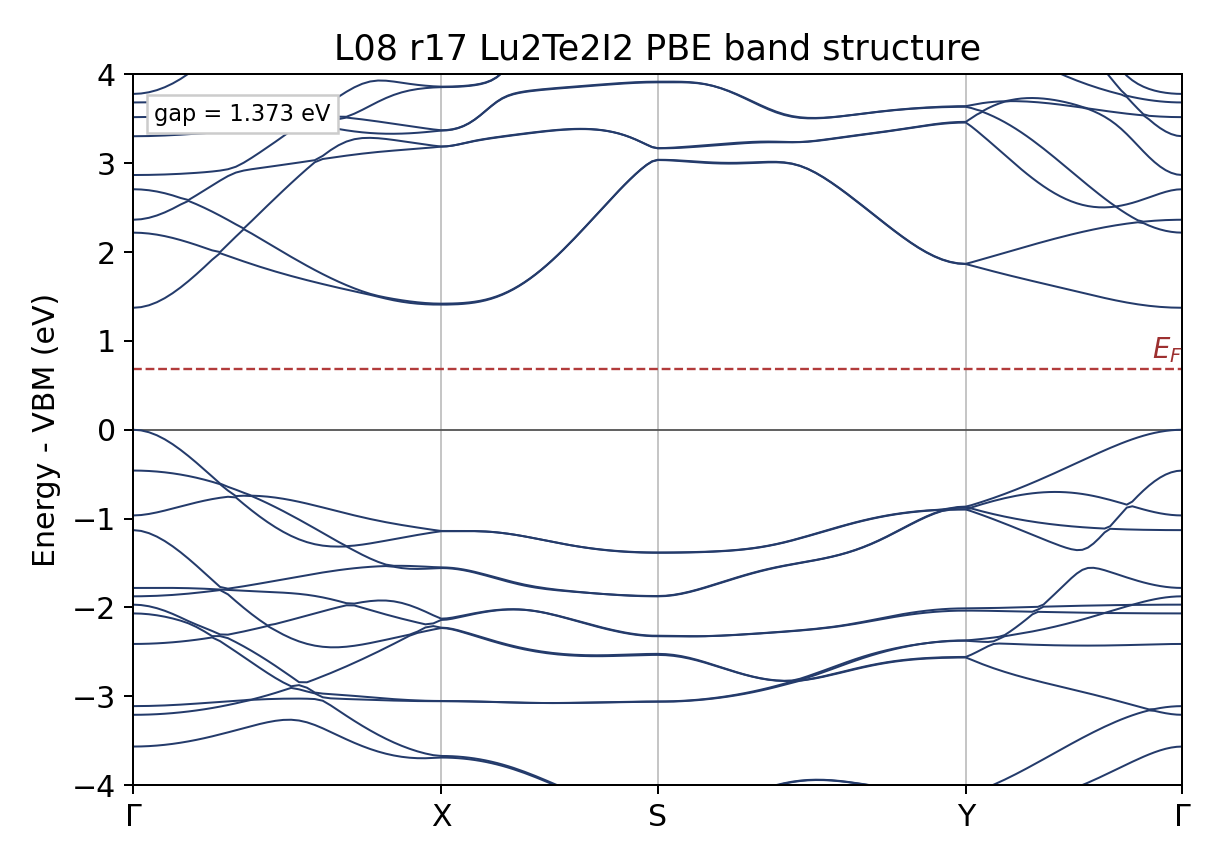} &
\includegraphics[width=0.30\textwidth]{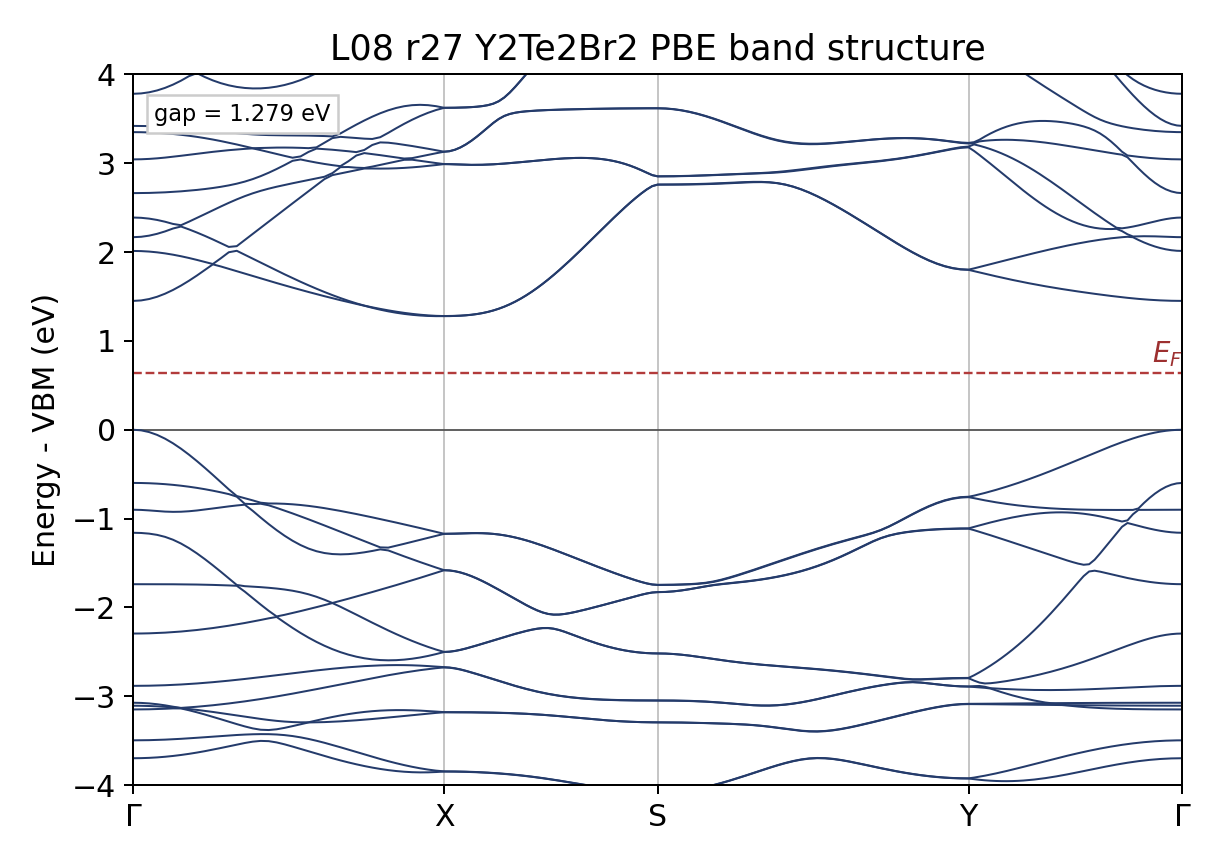} &
\includegraphics[width=0.30\textwidth]{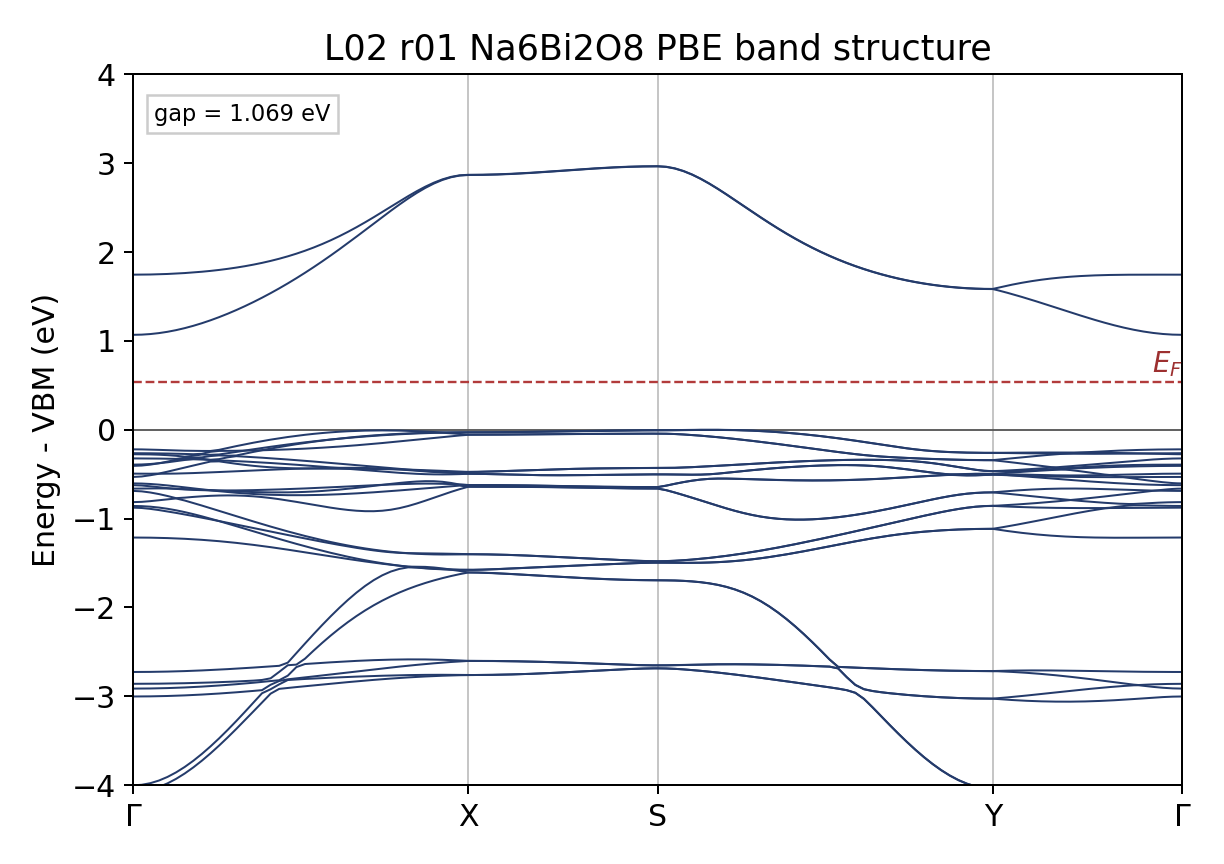} \\
\textbf{g} Ca$_2$Hg$_2$Br$_6$Cl$_2$ & \textbf{h} Ca$_2$Cd$_2$Br$_8$ & \textbf{i} Ca$_2$Hg$_2$Br$_8$ \\
\includegraphics[width=0.30\textwidth]{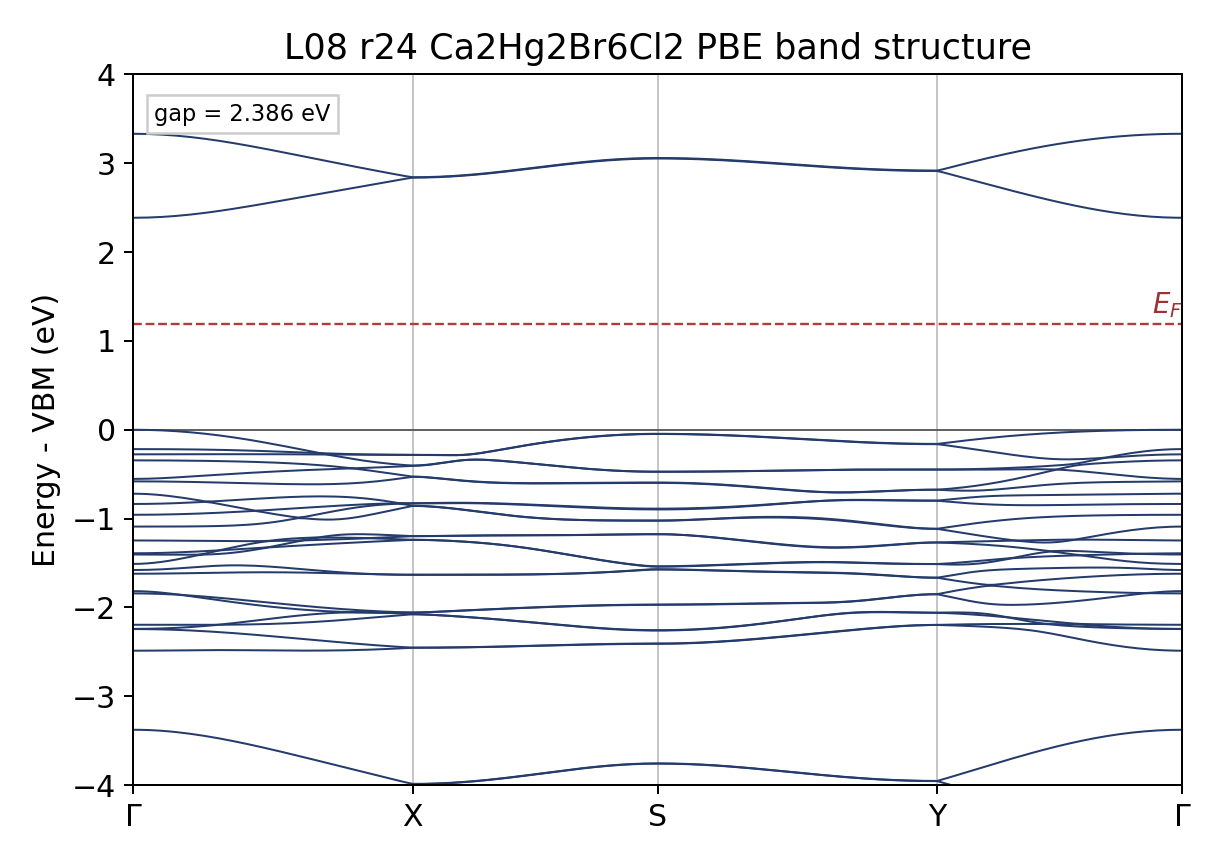} &
\includegraphics[width=0.30\textwidth]{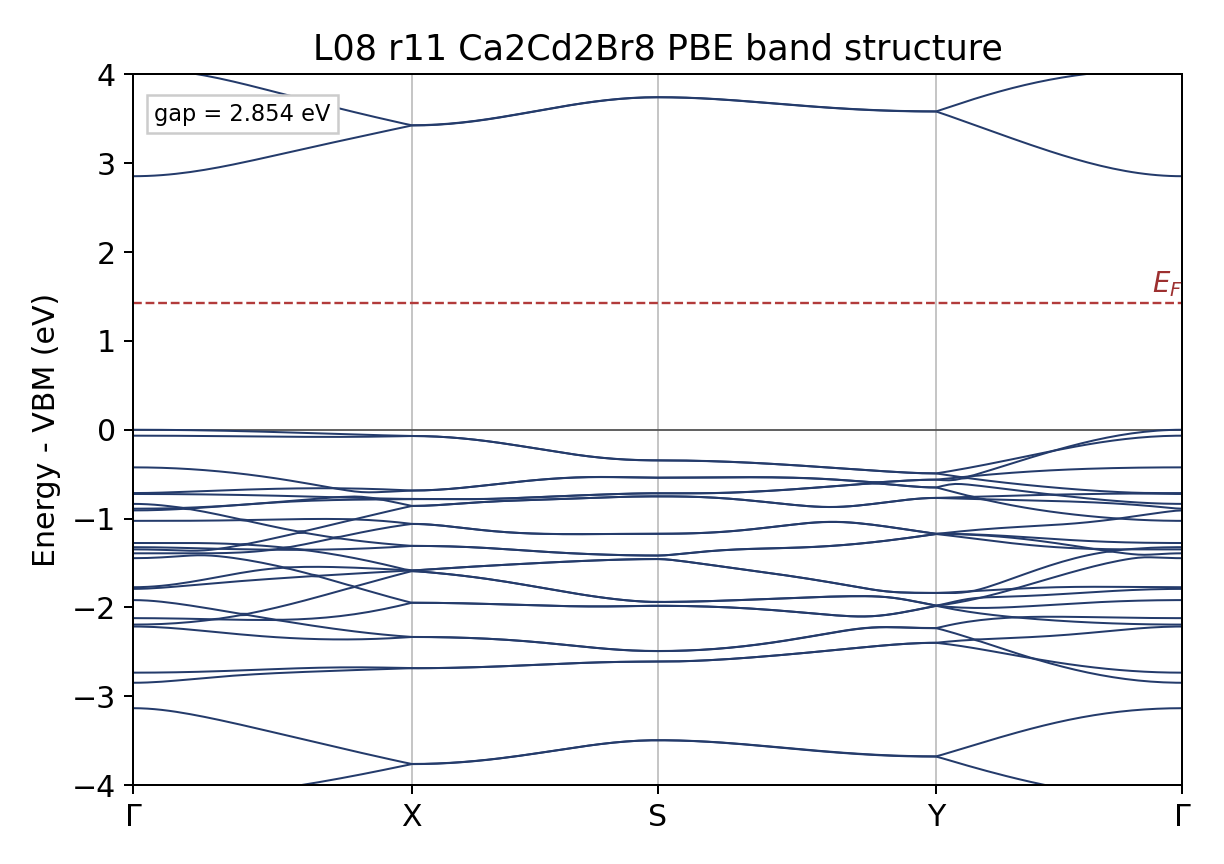} &
\includegraphics[width=0.30\textwidth]{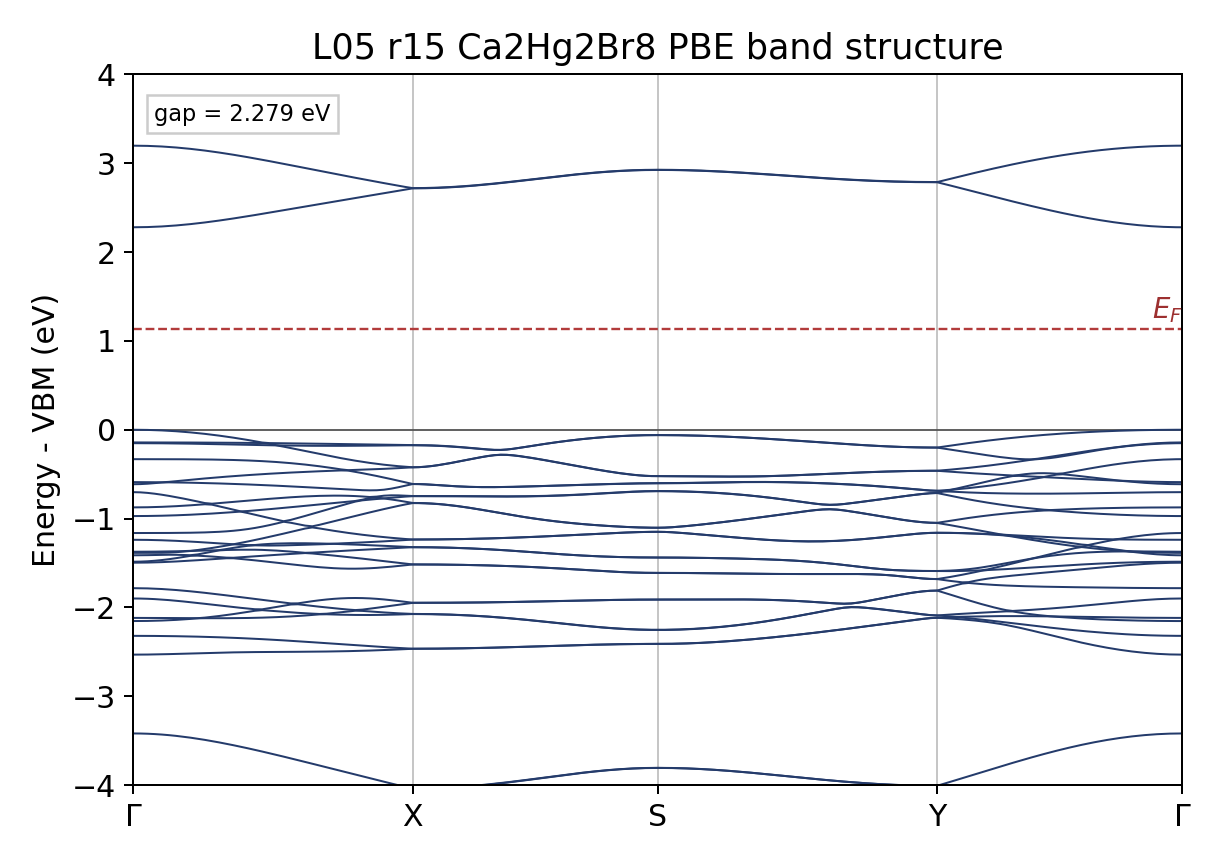} \\
\multicolumn{3}{c}{\textbf{j} Lu$_2$SeSCl$_2$} \\
\multicolumn{3}{c}{\includegraphics[width=0.30\textwidth]{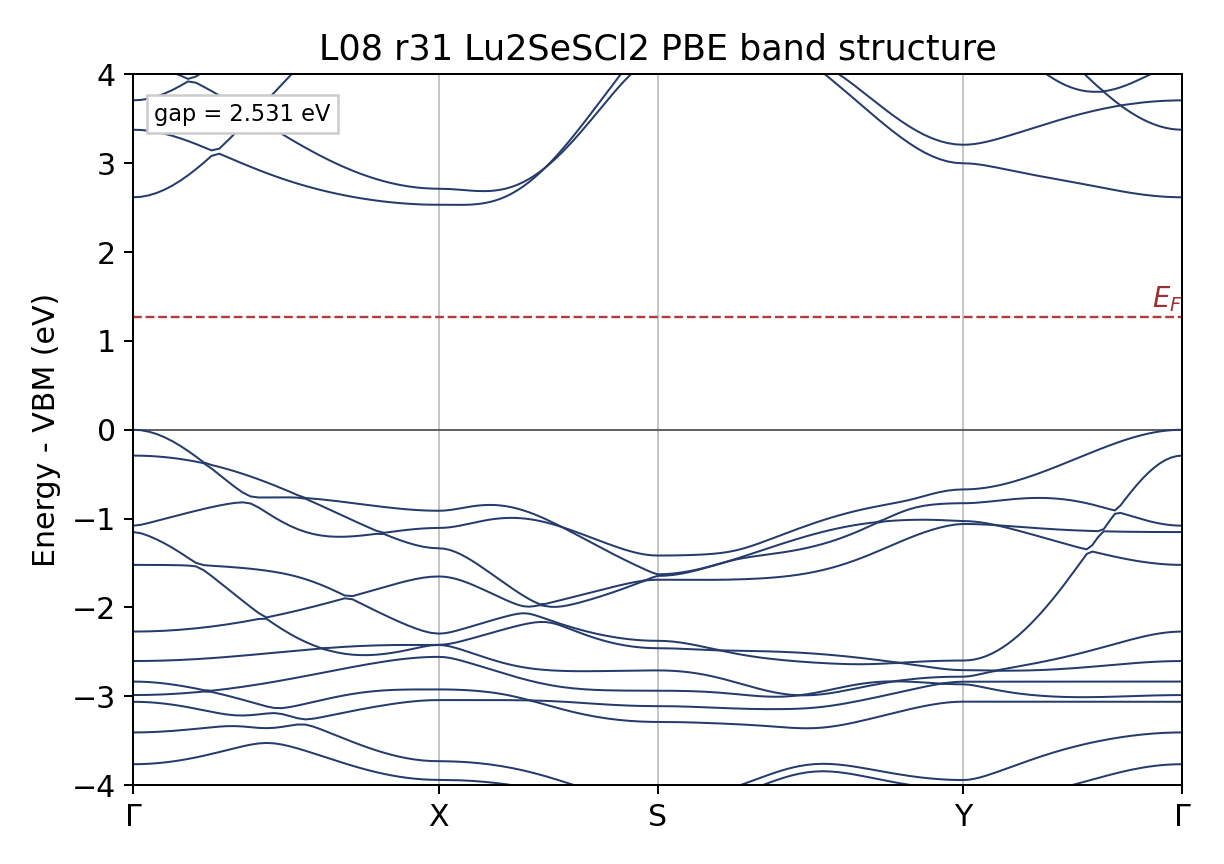}}
\end{tabular}
\caption{\textbf{Band structures of the ten highlighted generated formulas.} The panels show the band structures generated by the same multi-agent first-principles workflow used for the retained high-mobility evidence in Fig.~4.}
\label{fig:S5}
\end{figure}